\documentclass[authoryear,preprint,review,12pt]{elsarticle}

\RequirePackage[dvipsnames]{xcolor}
\usepackage[a4paper, top=1in, bottom=1in, left=1in, right=1in]{geometry} 
\usepackage{amssymb}
\usepackage{graphicx}
\usepackage{amsmath}
\usepackage{breqn}

\usepackage{placeins}
\usepackage{orcidlink} 
\usepackage{float}
\usepackage[dvipsnames]{xcolor}
\usepackage{silence}
\usepackage{subcaption}
\usepackage{booktabs}
\usepackage{tabularx}
\usepackage{multirow}

\DeclareUnicodeCharacter{2212}{-}

\journal{Journal of the Mechanics and Physics of Solids}

\begin{document}

\begin{frontmatter}



\title{Morphology-Property Interplay in Chemo-Mechanics of Ion-Intercalation Active Particles} 


\author{Rongyue Lin} 
\author{Royal C. Ihuaenyi} 
\author{Juner Zhu\corref{cor}}
\ead{j.zhu@northeastern.edu}
\author{Ruobing Bai\corref{cor}}
\ead{ru.bai@northeastern.edu}
\cortext[cor]{Corresponding authors.}

\affiliation{organization={Department of Mechanical and Industrial Engineering, Northeastern University},
            city={Boston},
            postcode={02115}, 
            state={MA},
            country={USA}}

\begin{abstract}
Morphology, material property, and mechanical constraint jointly govern the chemo-mechanical behavior of ion-intercalation particles, yet their coupled effects remain insufficiently understood. Here we establish a thermodynamically consistent single-particle framework and combine analytical solutions with multiphysics simulations to determine how these factors regulate lithiation and stress generation. We study hollow spherical, cylindrical, and ellipsoidal particles with isotropic or transversely isotropic material properties under fully constrained, inner-free, or unconstrained boundary conditions. We show that the transient lithiation pathway and the associated stress and strain fields are governed not by morphology, property, or constraint alone, but by their coupled interaction: isotropic particles are sensitive to the mechanical constraint, whereas transversely isotropic particles exhibit persistent heterogeneous lithiation dominated by anisotropic diffusivity. Flux decomposition analysis reveals substantial mechanical contributions to Li transport in both spherical and ellipsoidal particles, while increasing geometric anisotropy generally enhances the relative mechanical contribution and directional lithiation heterogeneity. Correlation analysis further shows that Li concentration and volumetric strain exhibit strong anti-correlation in unconstrained particles but weak correlation under full constraints. Bayesian optimization of hollow ellipsoids identifies Pareto-optimal morphologies that balance lithiation capacity against peak tensile stress. These results provide a unified framework for the morphology-property interplay in intercalation particles and offer morphology design principles for chemo-mechanical stability.
\end{abstract}




\end{frontmatter}



\section{Introduction}
\label{sec1}


Lithium-ion batteries (LIBs) have become the dominant electrochemical energy storage technology. As their applications expand from consumer electronics to more demanding use cases such as electric vehicles and grid-scale storage, the mechanics of electrodes has emerged as a critical determinant of the battery lifetime and performance. During a battery charge-discharge cycle, intercalation (lithiation) and deintercalation (delithiation) of Li ions induce considerable swelling and contraction of electrodes \citep{ipers_rapid_2024,oh_phenomenological_2016,hoschele_influence_2023}. Associated with this volumetric deformation is the complex evolution of the electrode microstructure \citep{song_microstructural_2025,Xu2021GuidingLearning,lu_multi-length_2021}, manifested by changes in characteristic features such as particle size, porosity, and tortuosity. Such microstructural evolution and deformation in electrodes can ultimately result in fracture and degradation of battery performance over repeated cycling \citep{edge_lithium_2021,han_review_2019,jiao_modeling_2026}.

Particle-scale chemo-mechanics is particularly important because heterogeneous Li distributions generate nonuniform transformation strains and internal stresses within active particles  \citep{Beaulieu2001ColossalAlloys,Schiele2018SiliconFracture, zhao_fracture_2010, Pharr2012KineticsBatteries}. The thermodynamics of coupled diffusion and deformation has been established through the classical works of \citet{prussin_generation_1961} and \citet{larche_linear_1973,larche_nonlinear_1978}, in which elastic deformation contributes to the chemical potential and thereby the mass transport. These principles have since been extensively applied to electrode materials and battery mechanics \citep{de_vasconcelos_chemomechanics_2022, zhao_review_2019}. Consequently, the fundamental constitutive ingredients of stress-dependent diffusion, chemical transformation strain, and elastic energy are well established. An important remaining question is instead how these mechanisms interact with realistic variations in particle morphology, material anisotropy, and mechanical confinement to regulate Li redistribution and stress localization.

Another key feature of electrode particles in both research and industrial applications is their diverse morphologies that depend strongly on the manufacturing process. Natural graphite particles are typically thin, plate-like flakes with high aspect ratios, while synthetic graphite particles are mostly spherical owing to production via spheroidization. In practice, graphite anode particles are often obtained by crushing or milling, yielding irregular shapes that are frequently highly porous \citep{zhu_testing_2018,Zhu2018InvestigationTests}. Cathode particles, such as lithium nickel manganese cobalt oxide Li(Ni$_x$Mn$_y$Co$_z$)O$_2$ (NMC) and lithium iron phosphate LiFePO$_4$ (LFP), generally appear as spherical or plate-like structures. A single secondary cathode particle typically consists of dozens of smaller primary particles bonded together by van der Waals forces \citep{han_microstructure_2025}. The morphology and crystallographic orientation of these primary particles are often engineered during manufacturing and sintering to achieve radial alignment. Moreover, secondary particles are commonly hollow, a feature that helps preserve structural integrity \citep{mao_highvoltage_2019,singh_microstructure-chemomechanics_2024,han_microstructure_2025,wang_reviving_2023}. Beyond commercial products, electrode particles can also be synthesized in rod-like \citep{zhang_rod-like_2013,xie_microsphere_2024} or other nonstandard geometries in the laboratory for research purposes, particularly to investigate directional Li diffusion and mechanical stabilization \citep{cai_critical_2016}.

The morphology of an electrode particle is intimately coupled to its material properties, including the elastic modulus, Li diffusivity, and lithiation/delithiation-induced spontaneous deformation, each of which may be isotropic or anisotropic. These couplings affect both the chemistry (e.g., Li intercalation) and the mechanics (e.g., internal stress and strain) of the electrode, with direct consequences for battery performance and stability during cycling. The couplings further raise a fundamental question: when a single particle incorporating a given amount of ions deforms, which particle morphology best promotes chemo-mechanical stability? \citet{christensen_stress_2006} introduced a stress-generation model for Li intercalation in single-phase spherical particles to study the influence of particle size and charge rate on fracture. Subsequent studies have found that layered cathode materials synthesized as near-equiaxed secondary particles can effectively reduce the initiation of microcracks by more uniformly distributing internal strain during cycling \citep{meng_morphology_2023}. Structural changes driven by anisotropic expansion of primary grains have also been observed, which can be suppressed by designing aligned nanoplate architectures~\citep{lee_development_2014,jiao_tuning_2022}. However, these existing efforts have each focused on specific morphologies (e.g., spherical or layered) and material properties (e.g., isotropy or anisotropy in modulus, diffusivity, and/or spontaneous deformation), leaving a systematic investigation of the morphology-property interplay in electrode chemo-mechanics largely unexplored.

A second knowledge gap concerns the lack of quantitative understanding of how mechanical constraints affect the chemo-mechanics of individual particles. In a real porous electrode containing millions of particles, each particle simultaneously deforms and interacts with neighboring particles, conductive and binder material (CBM), liquid electrolyte, and occasionally gas generated from side reactions during charge–discharge cycling. These complex interactions make it nearly impossible to identify the specific mechanical constraints acting on individual particles. Simulations based on reconstructed microstructures have revealed that the magnitude and distribution of stress fields are highly sensitive to particle arrangement, microstructural features, and spatial distribution of CBM \citep{xu_heterogeneous_2019,Xu2021GuidingLearning,lu_multi-length_2021,lu_3d_2020,song_microstructural_2025,hofmann_electro-chemo-mechanical_2020}. However, existing modeling approaches remain limited in their ability to reliably resolve particle-level mechanical constraints. To address this limitation, single-particle testing has emerged as a powerful electrochemical technique \citep{tsai_single-particle_2018,fang_fabrication_2022,fang_modeling_2021,song_digitaltwindriven_2023,lee_digitaltwindriven_2023}, in which an individual particle is isolated from a porous electrode, fabricated into a microelectrode, and cycled against a counter electrode. Nevertheless, the mechanical constraints present in such isolated testing differ significantly from those in real porous electrodes. Furthermore, recent experiments have shown that externally applied mechanical constraints can effectively extend battery lifetime by suppressing gas generation and delamination between component layers \citep{ihuaenyi_lifetime_2025}. Collectively, these observations underscore the critical role of mechanical constraints in determining electrode chemo-mechanics, yet their intrinsic coupling with particle morphology and material properties remains poorly understood.

In this work, we develop a coupled chemo-mechanical single-particle framework and apply it across three particle geometries, two classes of crystallographic material symmetry, and three representative types of mechanical constraints. Section 2 presents the formulation of the thermodynamically consistent theory. Section 3 applies the theory to spherical particles as a baseline, then to cylindrical particles where axial confinement and crystal orientation introduce additional coupling, and finally to ellipsoidal particles where the loss of high symmetry activates the full interplay among geometry, material anisotropy, and external constraint. Section 4 introduces two cross-field diagnostics, a projection-based flux decomposition to determine the signed contributions of chemical and mechanical driving forces to the resultant Li flux, and a Pearson correlation metric that quantifies the spatial correlation between Li concentration and deformation fields. Section 5 formulates a multi-objective optimization of geometric parameters and crystal orientation in hollow ellipsoidal particles to navigate the trade-off between electrochemical capacity and mechanical integrity.

\section{General Chemo-Mechanical Framework}
\label{sec:chemo-mechanical framework}

We start by developing a general chemo-mechanical framework based on a single secondary particle, a fundamental building block of a porous electrode. While the framework applies to both anode and cathode materials, this study will focus on ceramic-type cathode particles as a representative model system. Fig.~\ref{fig: system schematics} schematically illustrates the coupled chemical and mechanical processes considered in the framework. During discharge, the cathode particle undergoes Li-ion intercalation driven by the gradient of electrochemical potential. The intercalation-induced strain generates a stress field under certain mechanical constraints, which alters the total free energy of the system and thereby the diffusion kinetics. 

In this study, we choose a continuum framework for polycrystalline secondary particles, with a goal of using this framework to quantify how morphology and mechanical constraint govern the relative transport mechanisms, relationships among coupled physical fields, and morphology-performance trade-offs. Homogenized or effective-continuum particle models with idealized geometries have long been used to isolate diffusion-induced stress, chemical expansion, and stress-assisted transport without explicitly resolving the underlying polycrystalline microstructure \citep{ai_electrochemical_2020, cheng_evolution_2009, christensen_stress_2006, clerici_shape_2020}. Such models are complementary to grain-resolved descriptions, which are necessary when crystallographic incompatibility, grain-boundary stress localization, and intergranular fracture are the primary quantities of interest. Accordingly, the present model targets effective particle-scale transport and deformation rather than individual grain-boundary failure.

To investigate the morphology-property interplay in the electrode under different mechanical constraints, we formulate our framework based on the following mechanical and electrochemical assumptions.

First, from mechanics, we model the particle as a linear elastic material throughout Li intercalation. This assumption is justified because most transition-metal-layered oxides undergo small deformation (i.e., up to 5\% \citep{zhao_review_2019}) during intercalation, and intergranular fracture usually occurs before the onset of plastic deformation in a secondary agglomerate~\citep{yan_intragranular_2017}. While slip may still take place prior to fracture in single-crystal particles \citep{qian_single-crystal_2020,shu_unraveling_2025}, the current study focuses on particles made of oxide active materials that are mostly polycrystalline. 

Second, from electrochemistry, we model the particle as an ideal solution of Li ions and vacancies. The ideal solution model inherently assumes no phase separation or short-range ordering in the particle during intercalation, where all Li ions experience the same binding energy to the host lattice, independent of their local environment. For example, in a transition metal oxide (TMO$_2$), both the interlayer (sliding of O-TM-O slabs) and intralayer (hexagonal-monoclinic transformation) phase separations can be alleviated by operation at moderate charging/discharging rates \citep{liu_origin_2023} and proper doping and coating, respectively \citep{cho_licoosub_2001,shim_synergistic_2015,mikami_controlling_2024}. In the ideal solution model, Li ions and vacancies are assumed to be randomly mixed without correlated hopping, such that the enthalpy of mixing is neglected and only the entropy of mixing contributes to the free energy~\citep{zhao_phase_2022}. Furthermore, this model assumes an isopotential state within the particle, such that the electric potential is disregarded in the governing equation. The dielectric relaxation time of the particle is estimated as $\tau_e=\varepsilon_e/\sigma_e$, and the diffusion time of Li ions in the particle is estimated as $\tau_{\rm Li}=r^2/D$. Taking the permittivity $\varepsilon_e\approx10\varepsilon_0$ \citep{yasuhara_effect_2019}, the electronic conductivity $\sigma_e=10^{-4}$ S/m \citep{tukamoto_electronic_1997,kang_electrical_2019}, the characteristic length of particle $r=10$ $\mu$m, and the diffusivity of Li ions $D=10^{-14}$ m$^2$/s \citep{wiedemann_effects_2013}, we obtain $\tau_e\approx 10^{-6}$ s $\ll\ \tau_{\rm Li}\approx 10^4$ s. In other words, the system is limited by the diffusion of Li ions, while the electron transport reaches equilibrium in a much shorter time scale and can be regarded as isopotential.

The present formulation is intended for single-phase intercalation in small-strain elastic particles, with ideal Li-vacancy mixing and quasi-static mechanical equilibrium. Therefore, it isolates the coupling among transport, transformation strain, crystallographic anisotropy, and external constraint within a reduced but thermodynamically consistent setting. In addition, the formulation neglects the effect of grain boundaries when considering a polycrystalline secondary particle, such that a simplified thermodynamic framework with analytical or semi-analytical solutions can help highlight and quantify the morphology-property interplay under different mechanical constraints. Other effects not considered here, including phase separation, concentration-dependent material properties, plasticity, fracture, grain-boundary incompatibility, and explicit electrolyte/electronic overpotential fields, may alter quantitative predictions and, in some cases, qualitative behavior.

\subsection{Continuum framework} 

\begin{figure}[h!]
    \centering
    \includegraphics[width=0.9\linewidth]{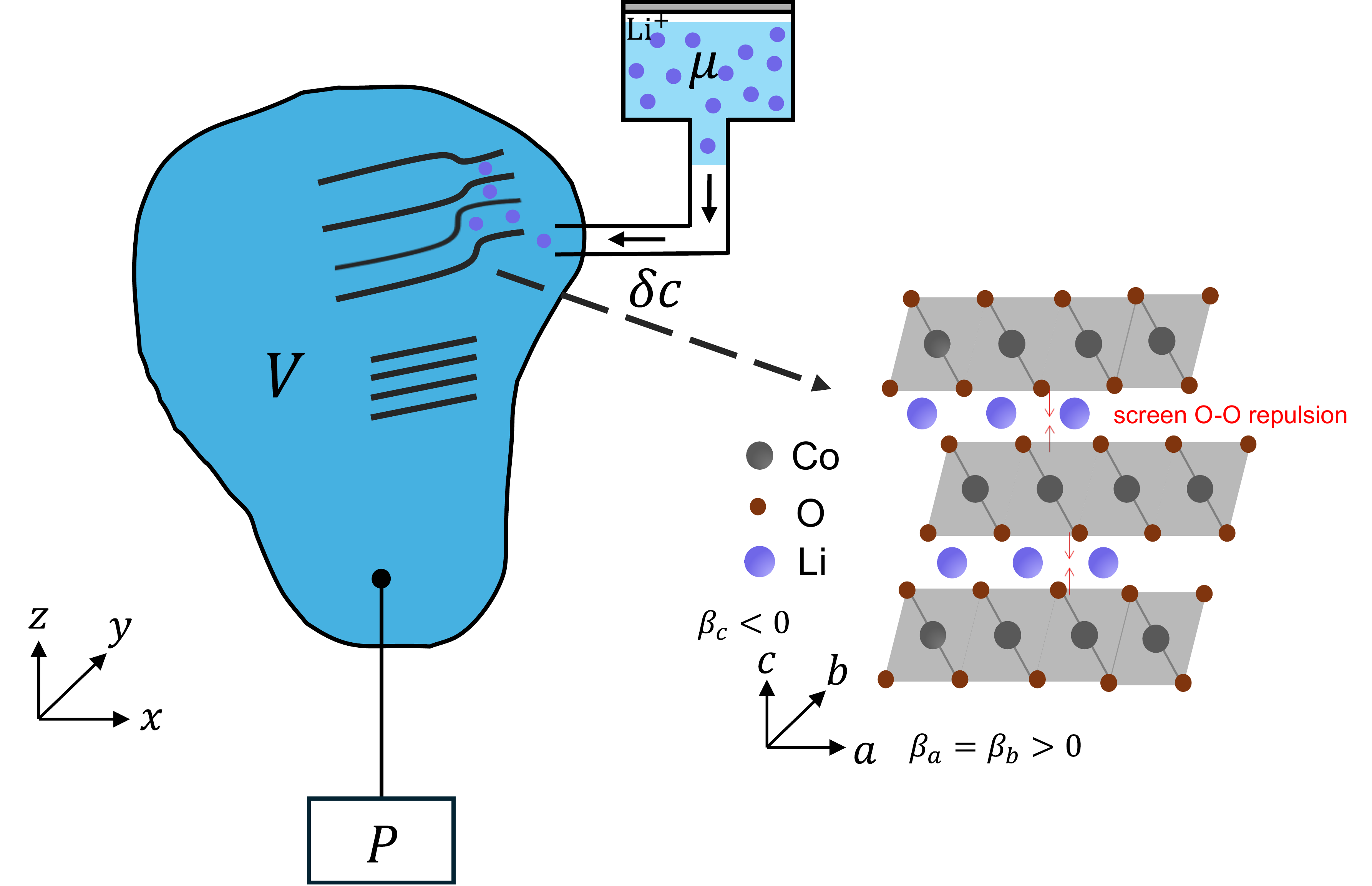}
    \caption{Schematic illustration of a representative electrode particle subjected to a mechanical load $\boldsymbol{P}$ and connected to a reservoir of Li ions with a prescribed chemical potential $\mu$. Li intercalation or deintercalation changes the local Li concentration and produces concentration-dependent lattice deformation. This lattice deformation, represented by the transformation strain, generates stress under mechanical constraints, which in turn contributes to the Li chemical potential and thereby alters the Li transport. The crystal-scale inset illustrates the anisotropic lattice deformation of layered LCO as a representative ceramic cathode material.}
    \label{fig: system schematics}
\end{figure}

Fig.~\ref{fig: system schematics} summarizes the physical picture underlying the continuum framework. At the particle scale, a representative electrode particle of volume $V$ exchanges Li ions with a reservoir prescribed by the Li chemical potential $\mu$, while being subjected to a mechanical load $\boldsymbol{P}$. Li insertion or extraction changes the local concentration \(c(\boldsymbol{x},t)\), where $\boldsymbol{x}$ is the 3D spatial coordinate and $t$ denotes the time elapsed since the onset of Li intercalation. This change of Li concentration generates a lattice deformation, represented by the concentration-dependent transformation strain. The crystal-scale inset illustrates this mechanism for layered LCO, where the transformation strain is anisotropic: Li insertion produces a small expansion in the basal plane while screening the electrostatic repulsion between adjacent oxygen layers, resulting in contraction along the crystallographic \(c\)-direction. The interplay between the transformation strain and mechanical constraints produces the stress field \(\boldsymbol{\sigma}(\boldsymbol{x},t)\) and strain field \(\boldsymbol{\varepsilon}(\boldsymbol{x},t)\), which subsequently enter the free energies and the governing equations for transport as shown below.

Under local equilibrium, associated with small variations $\delta\varepsilon_{ij}$ and $\delta c$, the Helmholtz free energy density changes by
\begin{equation} \label{helm energy}
    \delta w=\sigma_{ij}\ \delta\varepsilon_{ij}+\mu\delta c.
\end{equation}
Assuming $w=w(\boldsymbol{\varepsilon},c)$, we obtain the variation of free energy as
\begin{equation} \label{variation of helm energy}
\delta w
= \frac{\partial w(\boldsymbol{\varepsilon}, c)}{\partial \varepsilon_{ij}} \, \delta \varepsilon_{ij}
+ \frac{\partial w(\boldsymbol{\varepsilon}, c)}{\partial c} \, \delta c.
\end{equation}
Combining Eq.~\ref{helm energy} and Eq.~\ref{variation of helm energy}, we have
\begin{equation} \label{def stress}
    \sigma_{ij} = \frac{\partial w(\boldsymbol{\varepsilon}, c)}{\partial \varepsilon_{ij}},
\end{equation}
and
\begin{equation} \label{def chemical potential}
    \mu = \frac{\partial w(\boldsymbol{\varepsilon}, c)}{\partial c}.
\end{equation}

\subsection{Equilibrium and kinetics}
\label{equilibrium and kinetics}

The electrode particle, the external mechanical load, and the reservoir together form a thermodynamically closed system. In this system, the rate of change of the total free energy, ${\text{d} W}/{\text{d} t}$, is expressed as
\begin{eqnarray} \label{thermodynamics}
\frac{\text{d} W}{\text{d} t} 
&=& \int_V \frac{\partial w}{\partial t} \, \text{d}V - \int_A T_i \, \frac{\partial u_i}{\partial t} \, \text{d}A + \int_A {\mu J_i n_i \, \text{d}A},
\end{eqnarray}
where $T_i$ is the traction on the material boundary $A$, $u_i$ is the displacement field that satisfies $\varepsilon_{ij}=(u_{i,j}+u_{j,i})/2$, $J_i$ is the flux of the Li ions, and $n_i$ is the normal vector of the boundary. The three terms on the right hand side of Eq.~\ref{thermodynamics} represent changes in the bulk free energy of the particle, the potential energy of the external mechanical load, and the free energy of the reservoir, respectively. In addition, the conservation of Li ions in the particle follows
\begin{equation} \label{mass conservation}
    \frac{\partial c}{\partial t}=-\frac{\partial J_i}{\partial x_i}.
\end{equation}

Substituting Eq.~\ref{def stress} , Eq.~\ref{def chemical potential}, and Eq.~\ref{mass conservation} into Eq.~\ref{thermodynamics} and applying the divergence theorem, we obtain
\begin{eqnarray} \label{derive equilibrium}
\frac{\text{d} W}{\text{d} t} 
&=&  \int_A (\sigma_{ij} n_j - T_i) \frac{\partial u_i}{\partial t} \, \text{d}A - \int_V \frac{\partial \sigma_{ij}}{\partial x_j} \frac{\partial u_i}{\partial t} \, \text{d}V + \int_V \frac{\partial \mu}{\partial x_i} J_i \, \text{d}V.
\end{eqnarray}
The second law of thermodynamics dictates that ${\text{d} W}/{\text{d} t}\leq0$ for any arbitrary ${\partial u_i}/{\partial t}$ and ${\partial c}/{\partial t}$. To satisfy this condition, we assume mechanical equilibrium in the system, since the characteristic time of elastic deformation (governed by the speed of elastic wave) is much shorter than that of Li diffusion. As a result, the first two integrands in Eq.~\ref{derive equilibrium} vanish, leading to
\begin{equation} \label{BC}
    \sigma_{ij}n_j-T_i=0,
\end{equation}
as the traction boundary condition and
\begin{equation} \label{force balance}
    \frac{\partial\sigma_{ij}}{\partial x_j}=0,
\end{equation} 
as the balance of force in the bulk. Furthermore, we assume that the flux $\boldsymbol{J}$ is linearly proportional to the gradient of chemical potential $\partial\mu/{\partial x_j}$, following 
\begin{eqnarray} \label{def diffusion flux}
    J_i=-\mathrm{M}_{ij}c\frac{\partial\mu}{\partial x_j},
\end{eqnarray}
where $\boldsymbol{M}=\mathrm{M}_{ij}$ is a second-rank mobility tensor, related to the diffusivity tensor by the Nernst-Einstein relation $\mathrm{M}_{ij}=D_{ij}/(RT)$, and $RT$ is the thermal energy per mole of particles in unit of J/mol. So long as $\boldsymbol{M}$ (and $\boldsymbol{D}$) is positive-definite, we have ${\text{d} W}/{\text{d} t}\leq0$, satisfying the second law of thermodynamics.

\subsection{Free energies} 

We assume that the free energy $w(\boldsymbol{\varepsilon},c)$ can be additively decomposed into the elastic energy $w_{\rm elastic}(\boldsymbol{\varepsilon},c)$ and the chemical energy $w_{\rm chem}(c)$:
\begin{equation} \label{general free energy}
   w(\boldsymbol{\varepsilon},c)=w_{\rm elastic}(\boldsymbol{\varepsilon},c)+w_{\rm chem}(c).
\end{equation}
Assuming linear elasticity, we express the elastic energy as \citep{bai_photochemical-induced_2021,bai_collective_2021}
\begin{equation} \label{general strain energy}
   w_{\rm elastic}(\boldsymbol{\varepsilon},c)=\frac{1}{2}C_{ijkl}(\varepsilon_{ij}-\varepsilon_{ij}^{\rm chem}(c))(\varepsilon_{kl}-\varepsilon_{kl}^{\rm chem}(c)),
\end{equation}
where $C_{ijkl}$ is the fourth-rank stiffness tensor and $\boldsymbol{\varepsilon}^{\rm chem}(c)$ is the transformation strain depending on the Li concentration $c$. We assume the transformation strain $\boldsymbol{\varepsilon}^{\rm chem}(c)$ as a linear function of $c$: 
\begin{equation}\label{spontaneous def}
    \varepsilon_{ij}^{\rm chem}(c)=\beta_{ij}(c-c_0)=\beta_{ij}\tilde{c}, 
\end{equation}
where $\boldsymbol{\beta}=\beta_{ij}$ is the constant second-rank Vegard tensor, $c_0$ is the initial Li concentration in the particle, and $\tilde{c}=c-c_0$ is the amount of Li inserted. As a result, $\boldsymbol{\varepsilon}^{\rm chem}(c)$ is generalized beyond the common volumetric expansion or compression due to lithiation \citep{zhang_numerical_2007}. In particular, this generalized form of $\boldsymbol{\varepsilon}^{\rm chem}(c)$ captures the possible anisotropic deformation induced by lithiation, due to different electronic structures, binding energies, and crystal lattices \citep{moon_anisotropic_2019, zhou_fewlayer_2019, reimers_electrochemical_1992}. The linear form of $\varepsilon_{ij}^{\rm chem}(c)$ represents the equal contribution of the lithiation fraction to changes of all lattice parameters of the crystal \citep{malave_computational_2014}.

We express the chemical energy by assuming ideal mixing of Li ions and vacancies \citep{li_intrinsic_2021, shishvan_dendrites_2020, fang_modeling_2024}
\begin{eqnarray} \label{chemical free energy}
    w_{\rm chem}\left(c\right)
    = c\mu_{\rm Li}^0+\left(c_{\max}-c\right)\mu_V+c_{\max}RT\left(x_{\rm Li}\ln x_{\rm Li}+\left(1-x_{\rm Li}\right)\ln{\left(1-x_{\rm Li}\right)}\right),
\end{eqnarray}
where $\mu_{\rm Li}^0=-nFE_{\rm Li}^0$ is a constant that represents the chemical potential of Li in the stress-free reference state, $n=1$ is the number of electrons transferred per Li atom, $F$ is the Faraday constant, and $E_{\rm Li}^0$ is the standard electrode potential of Li with respect to the reference electrode. $\mu_V$ is the molar enthalpy of vacancy formation. $c_{\max}$ is a constant in unit of mol/m$^3$ representing the total number of intrinsic sites, such that $(c_{\max}-c)$ is the concentration of vacancies and $x_{\rm Li}={c}/c_{\rm max}$ represents the normalized Li composition.

\subsection{Material models} 

Combining Eq.~\ref{spontaneous def}, Eq.~\ref{chemical free energy}, Eq.~\ref{general free energy}, and Eq.~\ref{def stress}, the stress is expressed as
\begin{eqnarray} \label{constitutive eqn}
    \sigma_{ij}=C_{ijkl}\left(\varepsilon_{kl}-\varepsilon_{kl}^{\rm chem}\right)=C_{ijkl}\left(\varepsilon_{kl}-\beta_{kl}\tilde{c}\right).
\end{eqnarray}
Furthermore, combining Eq.~\ref{spontaneous def}, Eq.~\ref{general strain energy}, Eq.~\ref{general free energy}, and Eq.~\ref{def chemical potential}, the chemical potential is expressed as
\begin{eqnarray}
    \mu
    =\mu_0+RT\ln\frac{c}{c_{\max}-c}-C_{ijkl}\beta_{kl}\varepsilon_{ij}+C_{ijkl}\beta_{ij}\beta_{kl}\tilde{c},
\end{eqnarray}
where $\mu_0=\mu_{\rm Li}^0-\mu_V$ is a constant. Substituting this into Eq.~\ref{def diffusion flux}, the resultant flux is expressed as
\begin{equation} 
\label{general diffusion flux}
    J_i
    =-\frac{D_{ij}}{RT}(RT\frac{c_{\max}}{c_{\max}-c}\frac{\partial c}{\partial x_j}-C_{klmn}\beta_{mn} c \frac{\partial\varepsilon_{kl}}{\partial x_j}+C_{klmn}\beta_{kl}\beta_{mn}c\frac{\partial c}{\partial x_j}).
\end{equation}

\subsection{Material isotropy}
\label{material isotropy}

Atomic force microscopy (AFM) experiments have shown that particles made of amorphous active materials experience volumetric deformation with lithiation \citep{beaulieu_reaction_2003,timmons_isotropic_2007}. In this case, assuming material isotropy in all relevant properties of the particle, we can express the transformation strain, stiffness tensor, and diffusivity tensor as
\begin{equation} 
\label{isotropic transformation strain}
\varepsilon_{ij}^{\rm chem}\left(c\right)=\beta \tilde{c}\delta_{ij},
\end{equation}
\begin{equation} \label{eq:isotropic stiffness tensor}
C_{ijkl}=\lambda\delta_{ij}\delta_{kl}+G(\delta_{ik}\delta_{jl}+\delta_{il}\delta_{jk}),
\end{equation}
and
\begin{equation} 
D_{ij}=D\delta_{ij},
\end{equation}
where $\beta_{ij}=\beta\delta_{ij}$, $\delta_{ij}$ is the identity tensor, and $3\beta=\Omega$ is the partial molar volume of Li. $\lambda$ and $G$ are the Lame parameters related to the Young's modulus $E$ and the Poisson's ratio $\nu$ as 
\begin{equation} 
\lambda=\frac{E\nu}{(1+\nu)(1-2\nu)},
\end{equation}
\begin{equation} 
G=\frac{E}{2\left(1+\nu\right)},
\end{equation}
and $D$ is the isotropic diffusivity of Li in the material. With these, the stress tensor, chemical potential, and flux are simplified as
\begin{eqnarray}
\sigma_{ij}=\frac{E}{1+\nu}[(\varepsilon_{ij}-\beta \tilde{c}\delta_{ij})+\frac{\nu}{1-2\nu}(\varepsilon_{kk}-3\beta \tilde{c})\delta_{ij}],
\end{eqnarray}
\begin{eqnarray}
    \mu=\mu_0+RT\ln\frac{c}{c_{\max}-c}-\Omega\sigma_h,
\end{eqnarray}
and
\begin{equation} \label{isotropic flux}
    J_i=-\frac{D}{RT}\left[RT\frac{c_{\max}}{c_{\max}-c}\frac{\partial c}{\partial x_i}-c\Omega\frac{\partial \sigma_h}{\partial x_i}\right],
\end{equation}
where $\sigma_h=(\lambda+2G/3)(\varepsilon_{kk}-3\beta \tilde{c})$ is the hydrostatic stress.

\subsection{Model Parameters}
In this work, we take $\text{Li}_x\text{Co}\text{O}_2$ (LCO) as a model electrode particle that is widely used in portable devices because of its high volumetric energy density. We conduct simulations for three cases of different material properties: (1) isotropic with randomly oriented LCO crystals in spherical, cylindrical, and ellipsoidal particles, (2) anisotropic with LCO crystals aligned in the axial direction of a cylindrical particle, and (3) anisotropic with LCO crystals aligned in the radial direction of an ellipsoidal particle.

In our simulations, we adopt material parameters extracted from experimental data. For anisotropic material parameters, results from X-ray diffraction of single crystal shows significant anisotropic strains in the hexagonal lattice, such that at 50\% lithiation, the lattice parameter of the $a$ and $b$ axes has a small increase by 0.23\%, while the $c$ axis exhibits a strong contraction by approximately -2.39\%. The opposite signs of these lattice strains originate from the layered crystal structure: Li insertion produces a small in-plane expansion while screening the electrostatic repulsion between oxygen layers across the Li-containing interlayer, thereby reducing the interlayer spacing along the \(c\)-direction. These result in a volume change of around -1.91\% \citep{wang_effect_2005,reimers_electrochemical_1992}. Therefore, we take the crystallographically anisotropic Vegard coefficients in our model as $\beta_a=\beta_b=1.65\times10^{-6}$ $\mathrm{m^3/mol}$, and $\beta_c=-1.72\times10^{-5}$ $\mathrm{m^3/mol}$. In addition, we implement an anisotropic diffusivity tensor to simulate the Li flux between TMO$_2$ layers. All modeling parameters are summarized in Table \ref{table: modeling parameters}. Throughout the following numerical analyses, the normalized Li concentration is defined relative to the initial state as \((c-c_0)/(c_{\max}-c_0)\).

\begin{table}[h!]
\caption{Model parameters}
\label{table: modeling parameters}
\centering
\resizebox{\textwidth}{!}{
\begin{tabular}{lcc}
\toprule
\textbf{Parameter} & \textbf{Symbol} & \textbf{Value} \\
\midrule
Young's modulus \citep{mucke_modelling_2021}& $E$ & $123.6\ \mathrm{GPa}$ \\
Poisson's ratio \citep{mucke_modelling_2021}& $\nu$ & $0.288$ \\
\addlinespace
Elastic stiffness (transverse isotropy) \citep{yamakawa_phase-field_2018, wu_ab_2015}
& $C_{11}$ & $303.86\ \mathrm{GPa}$ \\
& $C_{12}$ & $101.71\ \mathrm{GPa}$ \\
& $C_{13}$ & $32.58\ \mathrm{GPa}$ \\
& $C_{33}$ & $98.93\ \mathrm{GPa}$ \\
& $C_{44}$ & $18.02\ \mathrm{GPa}$ \\
\addlinespace
Diffusivity \citep{wiedemann_effects_2013}& $D$ & $5.387\times 10^{-15}\ \mathrm{m^2/s}$ \\
Saturation Li concentration \citep{wiedemann_effects_2013} & $c_{\max}$ & $22.37\times 10^{3}\ \mathrm{mol/m^3}$ \\
Initial Li concentration \citep{wiedemann_effects_2013} & $c_0$ & $10.74\times 10^{3}\ \mathrm{mol/m^3}$ \\
\addlinespace
Isotropic Vegard coefficient& $\beta$ & $-4.6\times 10^{-6}\ \mathrm{m^3/mol}$ \\
\addlinespace
Anisotropic Vegard coefficient \citep{reimers_electrochemical_1992, malave_computational_2014}& $\beta_{a}$ & $1.65\times 10^{-6}\ \mathrm{m^3/mol}$ \\
& $\beta_{c}$ & $-1.72\times 10^{-5}\ \mathrm{m^3/mol}$ \\
\addlinespace
Partial molar volume& $\Omega$ & $-1.37\times 10^{-5}\ \mathrm{m^3/mol}$ \\
\addlinespace
Anisotropic diffusivity \citep{uxa_lithium-ion_2023}
& $D_a$ & $1.7\times 10^{-15}\ \mathrm{m^2/s}$ \\
& $D_c$ & $6\times 10^{-18}\ \mathrm{m^2/s}$ \\
\addlinespace
Temperature & $T$ & $298\ \mathrm{K}$ \\
\addlinespace
Discharging rate & & 1C \\
\bottomrule
\end{tabular}
}
\end{table}

\subsection{Computational Implementation}

The governing equations are solved using a combination of analytical reduction and finite element simulation, depending on the specific particle geometry. 

For spherical and cylindrical particles, geometric symmetry reduces the three-dimensional governing equations to one-dimensional problems in the radial coordinate. The mechanical equilibrium equations are solved analytically subject to the prescribed displacement or traction boundary conditions, yielding expressions for the displacement, stress, and strain fields as functions of the evolving Li concentration field. 

For ellipsoidal particles, because the geometry and loading considered here are rotationally symmetric about the long axis, the problem is formulated in the two-dimensional axisymmetric \(r\)-\(z\) domain using COMSOL Multiphysics 6.1. The constitutive expression for the Li flux is obtained directly from the chemical-potential formulation in Section 2 and, for arbitrarily oriented transversely isotropic materials, is given explicitly in \ref{app:analytical for ellips}. The governing equations for transient concentration and displacement fields are given in Section \ref{sec:ellip}.

Li transport is implemented using the Coefficient Form PDE interface, with \(c\) as the dependent variable and the chemo-mechanically coupled flux incorporated into the conservative transport equation. Mechanical equilibrium is implemented using the Solid Mechanics interface, with neglected inertial effects and the lithiation-induced transformation strain implemented through the External Strain feature as the Vegard tensor. The transient coupled problem is integrated using an implicit backward differentiation formula (BDF) with adaptive time stepping. A relative tolerance of \(10^{-4}\) is used for the time-dependent solver. A mesh-refinement study was performed for a representative ellipsoidal particle by comparing the COMSOL predefined finer, extra fine, and extremely fine meshes. The volume-averaged Li concentration and total elastic energy changed by less than 0.01\%, producing nearly overlapping responses over the entire lithiation range. These results demonstrate that further mesh refinement has a negligible effect on the quantities examined. The extra fine mesh was retained for the reported simulations as a conservative numerical choice.

For the morphology optimization, the finite-element model is coupled to the sequential model-based optimization workflow described in Section \ref{sec:opt}. Candidate geometric and crystallographic parameters are generated using the Tree-structured Parzen Estimator implemented in Optuna, and each candidate is passed to the finite-element model for evaluation of the prescribed electrochemical and mechanical objectives. The resulting objective values are subsequently returned to the optimizer to guide selection of the next candidate.

\section{Morphology-Property Interplay under Various Mechanical Constraints}
\label{sec:morphology-property int}

Building on the theoretical framework, we next use spherical, cylindrical, and ellipsoidal hollow particles as a controlled progression from high to reduced geometric symmetry to examine how morphology interacts with material anisotropy and mechanical constraint during lithiation. Rather than representing specific secondary-particle microstructures, these geometries enable the effects of shape and curvature to be systematically isolated. In addition, for each particle geometry, we study the effects of different boundary conditions that represent limiting states spanning from weak to strong mechanical constraints. We further consider isotropic and transversely isotropic stiffness, diffusivity, and transformation strain to quantify how crystallographic anisotropy modifies these morphology–constraint interactions. Within this setting, the calculations address particle-scale precursors to degradation, particularly lithiation heterogeneity, stress gradients, and tensile-stress localization, which can promote cracking, interfacial damage, and contact loss.

We describe each kind of mechanical constraint using an idealized set of boundary conditions. A particle with fully traction-free boundaries at both the outer and inner surfaces (herein called \textit{unconstrained}) represents an isolated or weakly constrained particle. A particle with both fixed-displacement surfaces (herein called \textit{fully constrained}) represents strong confinement by surrounding media, coating, or neighboring solids. A particle with a constrained outer surface but a traction-free inner surface (herein called \textit{inner-free}) represents a hollow particle only confined by surrounding media.

For chemical boundary conditions, we apply concentration-dependent insertion fluxes at the outer and inner surfaces of a hollow particle. The boundary conditions ensure a comparable charging across different particle geometries while preserving the leading site-availability dependence near the dilute and saturated limits. Specifically, we assume a steady-state Li flux at the outer surface as $-J|_{R}={j}/{F}$ and a zero flux at the inner surface. Here, $j=j_0g(\bar{c})$ with $g(\bar{c})=2\sqrt{\bar{c}(1-\bar{c})}$ and $\bar{c}=c/c_{\max}$, following \citep{allen_segregated_2021}. This choice retains the leading site-availability dependence of the Butler-Volmer exchange current in porous electrode theory \citep{doyle_modeling_1993}, for which $i_0\propto \sqrt{c(c_{\max}-c)}$ in the symmetric case. Accordingly, the flux vanishes in both the dilute and near-saturated limits and reaches its maximum in the range between. Factor 2 in $g(\bar{c})$ normalizes its value to the range of $[0,1]$, such that $j_0$ remains the nominal galvanostatic flux scale at 1C rate while $g(\bar{c})$ acts as a physically motivated throttling factor. Therefore, the boundary conditions are regarded as a reduced particle-level surrogate for the concentration dependence of intercalation kinetics, rather than a full Butler–Volmer law \citep{allen_segregated_2021}. This interpretation is also consistent with the coupled ion-electron theory (CIET), in which site occupancy and vacancy availability increasingly hinder insertion near full lithiation \citep{fraggedakis_theory_2021}.

\subsection{Spherical particles}\label{sec:spherical particles}

\begin{figure}[htpb]
    \centering
    \includegraphics[width=0.7\linewidth]{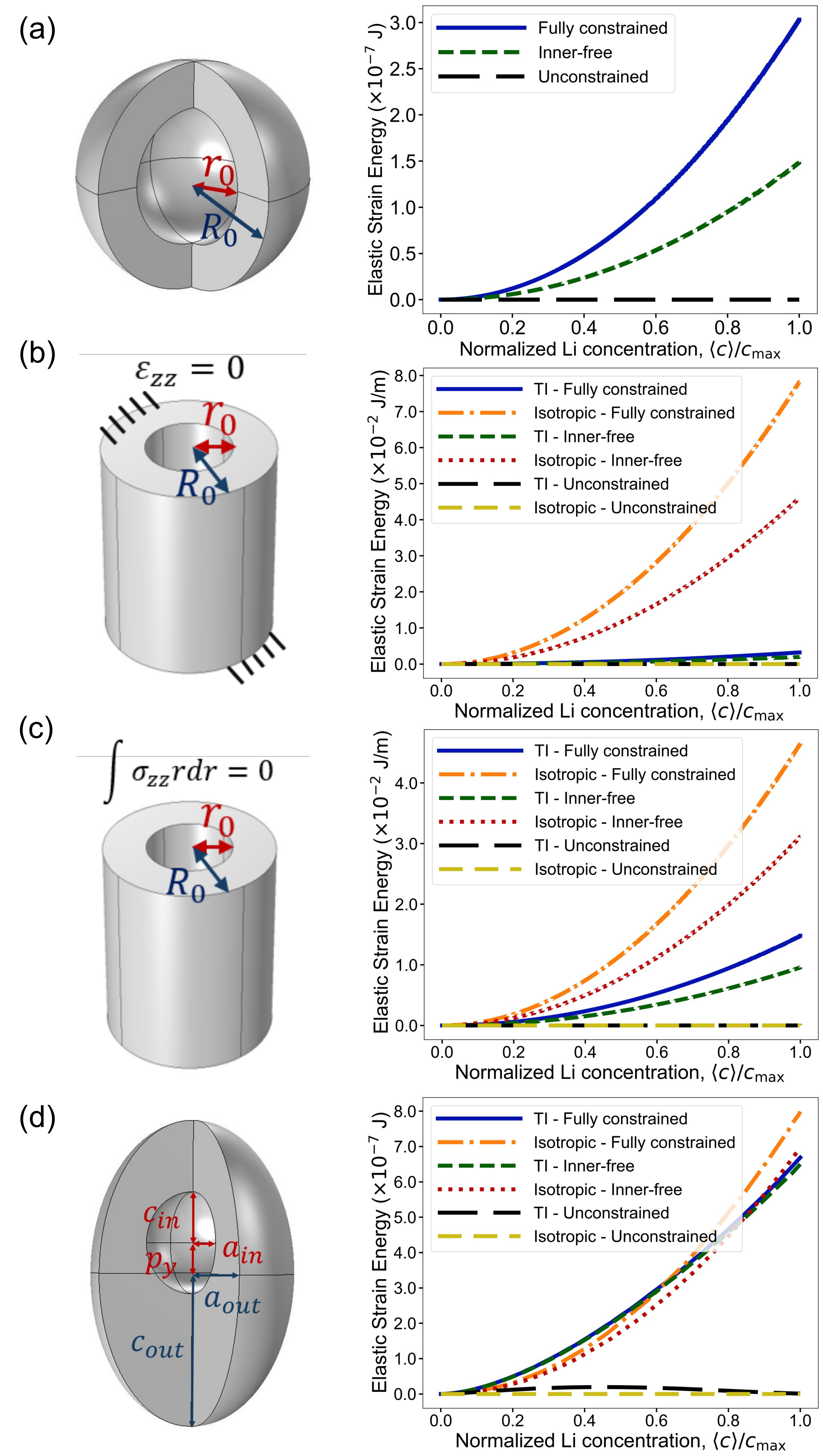}
    \caption{Evolution of elastic energy as a function of normalized volume-averaged Li concentration, $\langle c \rangle /c_{\max}$, for particles with different isotropic or transversely isotropic (TI) properties, morphologies, and mechanical constraints: fully constrained, inner-free, and unconstrained. (a) Spherical particle with isotropic properties. (b) Cylindrical particle with constrained axial displacement. (c) Cylindrical particle with unconstrained axial displacement (zero total axial force). (d) Ellipsoidal particle. In (a) and (d), the total elastic energy is plotted in unit of J. In (b) and (c), the elastic energy per unit axial length is plotted in unit of J/m.}
    \label{fig: energy comparison}
\end{figure}

Previous studies show that polycrystalline spherical particles with radially aligned single-crystal primary grains exposing ${010}$ facets enhance diffusion and cycling performance~\citep{xu_radially_2019}, whereas random grain orientations produce macroscopic isotropy~\citep{hu_evaluating_2025}.
We first study a hollow particle with isotropic material properties and spherical symmetry. We fix the ratio of inner and outer radii as $r_0/R_0=0.6$ (Fig.~\ref{fig: energy comparison}a left). 

The conservation of Li in Eq.~\ref{mass conservation} under spherical symmetry is
\begin{equation}
    \frac{\partial c}{\partial t} + \frac{1}{r^2} \frac{\partial (r^2J)}{\partial r}= 0,
\end{equation}
where the radial flux is
\begin{equation}
    J=-D_{\rm eff}\frac{\partial c}{\partial r} + \frac{Dc}{RT}(3\lambda+2G)\beta \frac{\partial}{\partial r}(\frac{\partial u_r}{\partial r}+2\frac{u_r}{r}),
\end{equation}
with the effective diffusivity
\begin{equation}
D_{\rm eff} = D\frac{c_{\max}}{c_{\max}-c} + \frac{Dc}{RT}\left(9\lambda+6G\right)\beta^2,
\end{equation}
and the radial displacement $u_r$.

The balance of force in Eq.~\ref{force balance} is simplified as
\begin{equation} \label{spherical force balance}
    \frac{\partial \sigma_r}{\partial r}+\frac{2}{r}\left(\sigma_r-\sigma_\theta\right)=0,
\end{equation}
where the radial and hoop stresses are obtained from Eq.~\ref{constitutive eqn} as
\begin{equation}
    \sigma_r=2G\left(\frac{\partial u_r}{\partial r}-\varepsilon_r^{\rm chem}\right)+\lambda\left[\left(\frac{\partial u_r}{\partial r}+2\frac{u_r}{r}\right)-\left(\varepsilon_r^{\rm chem}+2\varepsilon_\theta^{\rm chem}\right)\right],
\end{equation}
\begin{equation}
    \sigma_\theta=2G\left(\frac{u_r}{r}-\varepsilon_\theta^{\rm chem}\right)+\lambda\left[\left(\frac{\partial u_r}{\partial r}+2\frac{u_r}{r}\right)-\left(\varepsilon_r^{\rm chem}+2\varepsilon_\theta^{\rm chem}\right)\right].
\end{equation}
Here, $\varepsilon_r={\partial u_r}/{\partial r}$ and $\varepsilon_{\theta}={u_r}/{r}$ are the radial and hoop strains. $\varepsilon_r^{\rm chem}=\varepsilon_{\theta}^{\rm chem}=(\Omega/3) \tilde{c}$ are the isotropic transformation strains. Combining the above equations and solving the ODE analytically, we obtain
\begin{equation}\label{u_r spherical}
    u_r\left(r\right)=\frac{1+\nu}{1-\nu}\frac{\Omega}{3}\frac{1}{r^2}\ \int_{r_0}^{r}{\widetilde{c}r^2\text{d}r}+Ar+\frac{B}{r^2},
\end{equation}
\begin{equation}\label{sigma_r spherical}
    \sigma_r(r)=E\left[-\frac{2}{1-\upsilon}\frac{\Omega}{9}\frac{3}{r^3}\ \int_{r_0}^{r}{\widetilde{c}r^2\text{d}r}+\frac{A}{1-2\upsilon}-\frac{2B}{r^3\left(1+\upsilon\right)}\ \right],
\end{equation}
\begin{equation} \label{sigma_theta spherical}
    \sigma_\theta(r)=E\left[\frac{1}{1-\upsilon}\frac{\Omega}{9}\frac{3}{r^3}\ \int_{r_0}^{r}{\widetilde{c}r^2\text{d}r}-\frac{1}{1-\upsilon}\ \frac{\Omega}{3}\widetilde{c}\ +\frac{A}{1-2\upsilon}+\frac{B}{r^3\left(1+\upsilon\right)}\ \right].
\end{equation}
The constants $A$ and $B$ are determined from specific boundary conditions: fully constrained ($u_r|_{r=r_0, R_0}=0$), inner-free ($\left.\sigma_r\right|_{r=r_0}=0$ and $\left.u_r\right|_{r=R_0}=0$), and unconstrained ($\left.\sigma_r\right|_{r=r_0}=0$ and $\left.\sigma_r\right|_{r=R_0}=0$).

Using material parameters from Table \ref{table: modeling parameters}, we model the discharge of the hollow spherical particle until full lithiation ($t=3600$ s). We plot the total elastic energy as a function of the normalized Li concentration averaged over the entire particle, $\langle c \rangle / c_{\max}$, for particles under different mechanical constraints: fully constrained, inner-free, and unconstrained (Fig.~\ref{fig: energy comparison}a right). During discharge, the elastic energy increases monotonically in fully constrained and inner-free particles, the former more dramatically than the latter. By contrast, the elastic energy maintains nearly zero in unconstrained particles, indicating negligible lithiation-induced elasticity. 
These trends in elastic energy under different mechanical constraints have been widely reported in literature, showing the direct consequence of lithiation-induced transformation strain, $(\Omega/3)\tilde{c}$ \citep{christensen_stress_2006, de_vasconcelos_chemomechanics_2022}.


We also analytically compare the mechanical contribution (from the lithiation-induced transformation strain) and the chemical contribution (from the classical Fickian diffusion-reaction) to Li transport. Combining Eqs. \ref{def chemical potential} and \ref{general free energy}, we obtain the mechanical contribution to the Li chemical potential as ${\mu }_{\mathrm{m}}=\partial w_{\mathrm{elastic}}/\partial c$ and the chemical contribution as ${\mu }_{\mathrm{c}}=\partial w_{\mathrm{chem}}/\partial c$. In the case of an isotropic particle with spherical symmetry, substituting these expressions together with Eq. \ref{u_r spherical} into Eq. \ref{general diffusion flux}, we obtain the mechanical-driven and chemical-driven radial fluxes as
\begin{equation}
    J_{\mathrm{m}}=-\frac{2DE{\beta }^{2}c}{RT\left(1-\nu \right)}\frac{\partial c}{\partial r},
\end{equation}
and
\begin{equation}
    J_{c}=-D\frac{c_{\max}}{c_{\max}-c}\frac{\partial c}{\partial r},
\end{equation}
respectively. The ratio between the two fluxes is
\begin{equation}
    \frac{J_{\mathrm{m}}}{J_{\mathrm{c}}}=\frac{2E{\beta }^{2}c_{\max}}{RT\left(1-\nu \right)}x_{\rm{Li}}\left(1-x_{\rm{Li}}\right),
    \label{eq:Jm/Jc}
\end{equation}
in terms of the normalized local Li fraction $x_{\mathrm{Li}}=c/c_{\max}$. Substituting the isotropic material parameters from Table 1 into Eq.~\ref{eq:Jm/Jc} gives \(J_{\mathrm{m}}/J_{\mathrm{c}} \approx 16.6\) at \(x_{\mathrm{Li}}=0.5\), but \(J_{\mathrm{m}}/J_{\mathrm{c}} \approx 0.66\) at \(x_{\mathrm{Li}}=0.99\). This estimate suggests that the mechanical contribution substantially exceeds the chemical contribution during the lithiation but becomes secondary near full lithiation. The dominant mechanical contribution during lithiation is likely attributed to the nonuniform hydrostatic stress and associated mechanical driving forces induced by the gradient of radial concentration.

This theoretical analysis is confirmed by the quantitative results shown in Fig.~\ref{app fig:c profile} (Appendix A). We compare the fully coupled model with a concentration-driven model (i.e., a Fickian-like process without including the mechanical contribution to the chemical potential). The radial concentration profiles from the two solutions remain close throughout lithiation, as shown at 3 representative lithiation states, $\left<c\right>/c_{\max}=$ 0.25, 0.5, and 0.75 (Figs.~\ref{app fig:c profile}a-c). At the intermediate state, \(\left<c\right>/c_{\max}=0.5\) (Fig.~\ref{app fig:c profile}d), the mechanical flux \(J_{\mathrm{m}}\) substantially exceeds the chemical flux \(J_{\mathrm{c}}\). Consistently, as will be introduced in Section 4.1, the mechanical global flux ratio \(C_{\mathrm{m}}\) remains around 0.9, dominating over the chemical global flux ratio \(C_{\mathrm{c}}\) over much of lithiation (Fig.~\ref{app fig:c profile}e). However, as lithiation nearly completes, \(C_{\mathrm{m}}\) dramatically decreases to around 0.4 and becomes secondary to the chemical contribution \(C_{\mathrm{c}}\). These results demonstrate that similar concentration profiles do not necessarily imply a negligible mechanical contribution to Li transport. Rather, the spherical particle exhibits substantial chemo-mechanical coupling, whose relative importance evolves with lithiation.

\subsection{Cylindrical particles}
Next, we investigate cylindrical morphology as another common form of electrode particles (Fig.~\ref{fig: energy comparison}b\&c). Compared to spherical particles, cylindrical particles can exhibit additional failure mechanisms, such as significant increase in length during lithiation \citep{chan_high-performance_2008} and buckling under large axial constraints by a stiff coating material \citep{chakraborty_combining_2015, zang_silicon_2012}. The chemo-mechanics of a cylindrical particle may further depend on its possible anisotropy, for example, crystals can align along the axial direction of the cylinder, leading to transversely isotropic properties.

We model a hollow cylindrical particle with the same ratio of inner and outer radii as $r_0/R_0=0.6$. We assume the length of cylinder is much larger than its radius $R_0$. Under cylindrical symmetry, all stress, strain, and concentration fields depend only on the radial coordinate $r$. The conservation of Li and the balance of force are simplified as
\begin{equation}
    \frac{\partial c}{\partial t} + \frac{1}{r} \frac{\partial (rJ_r)}{\partial r}= 0,
\end{equation}
and
\begin{equation}
    \frac{\partial \sigma_{rr}}{\partial r}+\frac{\sigma_{rr}-\sigma_{\theta\theta}}{r}=0,
\end{equation}
respectively, where $J_r$ is the flux in the radial direction, and $\sigma_{rr}$ and $\sigma_{\theta\theta}$ are the radial and hoop stresses. In the axial direction, we assume that the particle is either fully constrained in displacement (axial strain $\varepsilon_{zz}=0$, Fig.~\ref{fig: energy comparison}b left) or unconstrained (total axial force $\int_{r_0}^{R_0} \sigma_{zz}r\text{d}r = 0$, Fig.~\ref{fig: energy comparison}c left). See further detailed analytical expressions in \ref{app:analytical for cylinders}.

We compare two types of material properties that are commonly observed in cylindrical particles. An isotropic particle follows the same description of diffusivity, transformation strain, and elastic constants as discussed in Section \ref{material isotropy}. A transversely isotropic particle has the symmetry axis $c$ of the crystal lattice aligned with the cylindrical axial axis $z$, such that the diffusivity tensor is expressed as $\boldsymbol{D}=\operatorname{diag}(D_a, D_a, D_c)$, the Vegard tensor $\boldsymbol{\beta}=\operatorname{diag}(\beta_a, \beta_a, \beta_c)$, and the stiffness tensor $\boldsymbol{C}$ composed of five independent constants $C_{11}, C_{12}, C_{13}, C_{33}, C_{44}$, with $C_{66}=(C_{11}-C_{12})/2$.

Figs.~\ref{fig: energy comparison}b\&c show the evolution of elastic energy per unit axial length in isotropic or transversely isotropic cylindrical particles during discharge, under various mechanical constraints. Similar to the case of spherical particle, lithiation-induced increase in elastic energy depends strongly on mechanical constraints: the most significant increase in fully constrained particles, less so in inner-free particles, and negligible in unconstrained particles. The axial constraint further amplifies this effect (i.e., comparing Figs.~\ref{fig: energy comparison}b and c).

Table~\ref{tab:cylinder_axial_stress_strain} summarizes the axial stress and strain in isotropic and transversely isotropic particles at the end of discharge under different radial mechanical constraints. When being axially unconstrained (Table~\ref{tab:cylinder_axial_stress_strain} bottom row), both particles experience lithiation-induced contraction in the axial direction due to their negative Vegard coefficients. However, in the radial and hoop directions, while the isotropic particle undergoes the same contraction, the transversely isotropic particle expands due to the positive Vegard coefficient $\beta_a=1.65\times 10^{-6}\ \mathrm{m^3/mol}>0$. The opposite signs of the axial and basal-plane transformation strains partially accommodate the lithiation-induced deformation and thereby reduce the elastic energy stored in the transversely isotropic particle. Consequently, under the same constraints, less elastic energy builds up compared to that in the isotropic particle (comparing curves in Figs.~\ref{fig: energy comparison}b\&c). When the particles are axially constrained, suppression of the lithiation-induced axial contraction generates tensile axial stress (Table~\ref{tab:cylinder_axial_stress_strain} top row). The stress levels are similar between two particles and decrease with fewer mechanical constraints (i.e., from fully constrained, inner-free, to unconstrained). 

\begin{table}[h!]
\caption{Axial stress and strain under different radial mechanical constraints at the end of discharge}
\label{tab:cylinder_axial_stress_strain}
\centering
\renewcommand{\arraystretch}{1.2}
\begin{tabularx}{\linewidth}{llXXX}
\toprule
 & & \textbf{Fully constrained} & \textbf{Inner-free} & \textbf{Unconstrained} \\
\midrule
\multirow{2}{*}{\shortstack{Axial stress [MPa] \\(axially constrained)}}
 & Isotropic & 17.4 & 11.6 & 0.73 \\
 & TI        & 27.3 & 23.7 & 2.32 \\
 \midrule
\multirow{2}{*}{\shortstack{Axial strain\\(axially unconstrained)}}
 & Isotropic & -0.011 & -0.0084 & -0.0059 \\
 & TI        & -0.0209 & -0.0215 & -0.0223 \\
\bottomrule
\end{tabularx}
\end{table}

\subsection{Ellipsoidal particles} \label{sec:ellip}

We further study a hollow ellipsoidal particle that represents a more generalized case of irregular, spherical-like morphology. As illustrated in Fig.~\ref{fig: energy comparison}d left, the ellipsoid contains a spherical inner cavity, with its position described by geometric parameters, $a_{\text{in}}=c_{\text{in}}=2$ $\mu$m and $p_y=0$. The aspect ratio of the ellipsoid is set as $a_{\text{out}}/c_{\text{out}}=0.5$. We use cylindrical coordinates $(r,\theta,z)$ and assume rotational symmetry about the long axis $z$, such that all fields are independent of $\theta$. Later in Section~\ref{sec:opt}, we will take this configuration as a baseline case and treat all geometric parameters as design variables for an optimization study. 

With these simplifications, the conservation of Li is expressed as
\begin{equation}
    \frac{\partial c}{\partial t} + \frac{1}{r} \frac{\partial (rJ_r)}{\partial r} + \frac{\partial J_z}{\partial z} = 0.
\end{equation}
The balance of force is governed by
\begin{eqnarray}
    \frac{\partial \sigma_{rr}}{\partial r}+\frac{\sigma_{rr}-\sigma_{\theta\theta}}{r}+\frac{\partial \sigma_{rz}}{\partial z}=0, \\ \nonumber
    \frac{\partial \sigma_{rz}}{\partial r}+\frac{\sigma_{rz}}{r}+\frac{\partial \sigma_{zz}}{\partial z}=0.
\end{eqnarray}
 
For the ellipsoidal morphology, we consider an isotropic particle using the same description as discussed in Section \ref{material isotropy}, and further consider a transversely isotropic particle with the crystal \textit{c}-axis aligned along the radial direction $r$ and the crystal \textit{a-b} plane coinciding with the $\theta$-$z$ plane. Consequently, the diffusivity tensor is $\boldsymbol{D}=\operatorname{diag}(D_c, D_a, D_a)$, the Vegard tensor is $\boldsymbol{\beta}=\operatorname{diag}(\beta_c, \beta_a, \beta_a)$, and the stiffness tensor $\boldsymbol{C}$ is composed of five independent constants $C_{11}, C_{12}, C_{13}, C_{33}, C_{44}$. In each case of different material properties and mechanical constraints, we calculate the flux analytically, as detailed in \ref{app:analytical for ellips}, and then solve the differential equations for stress, strain, and displacement fields numerically using COMSOL Multiphysics (Solid Mechanics Module). Later in Section~\ref{sec:opt}, we will also take this radially aligned crystal configuration as a baseline case and treat the crystal alignment as a design parameter. 

We first examine the evolution of total elastic energy during lithiation (Fig.~\ref{fig: energy comparison}d). Similar to the spherical and cylindrical cases, elastic energy in ellipsoidal particles increases monotonically with lithiation under fully constrained and inner-free conditions but remains near-zero when unconstrained. However, there is no significant difference between the former two mechanical constraints or between the isotropic and transversely isotropic particles. These observations indicate that the ellipsoidal morphology itself plays a dominant role in determining the total lithiation-induced elastic energy. 

\begin{figure}[htbp]
    \centering
    \includegraphics[width=1\linewidth]{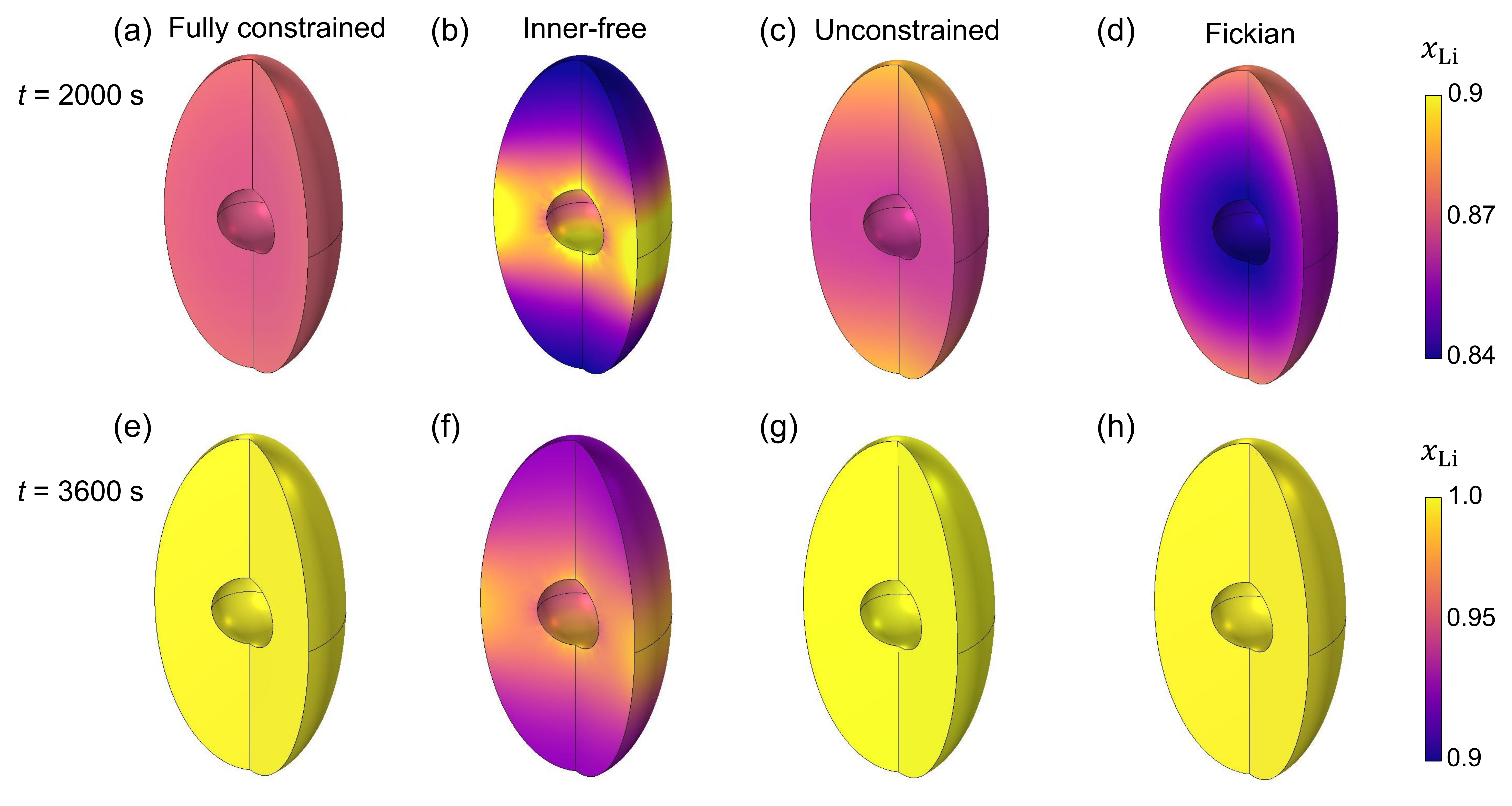}
    \caption{Full fields of normalized Li concentration $x_{\rm Li}=c/c_{\rm max}$ in an isotropic ellipsoidal particle at $t=2000$ s (first row) and $t=3600$ s (second row) under different mechanical constraints: (a)\&(e) fully constrained, (b)\&(f) inner-free, (c)\&(g) unconstrained, and (d,h) Fickian-like process neglecting elasticity.}
    \label{fig: isotropic ellipsoid}
\end{figure}

To further examine the effects of mechanical constraints and material properties, we calculate the full field of Li concentration, $x_{\text{Li}}=c/c_{\text{max}}$, in the ellipsoidal particle at both an intermediate stage ($t=2000$ s) and the end of discharge ($t=3600$ s) under 1C. In an isotropic particle at $t=2000$ s, the concentration field depends strongly on the mechanical constraint. Fully constrained boundaries lead to a near-homogeneous lithiation of $x_{\text{Li}} \approx 0.88$ (Fig.~\ref{fig: isotropic ellipsoid}a). Unconstrained boundaries introduce slight heterogeneity and anisotropy (Fig.~\ref{fig: isotropic ellipsoid}c). In contrast, inner-free boundaries produce the most pronounced heterogeneous and anisotropic lithiation (Fig.~\ref{fig: isotropic ellipsoid}b), despite the isotropy of the material itself: lithiation proceeds fastest along the short axis and slowest along the long axis, with markedly different Li concentrations of $x_{\text{Li}} \approx 0.9$ and $x_{\text{Li}} \approx 0.84$, respectively. Furthermore, a diffusion-reaction process that neglects elasticity (i.e., $w_{\rm elastic}(\boldsymbol{\varepsilon},c)$ in Eq.~\ref{general strain energy}) yields heterogeneous but nearly isotropic lithiation (Fig.~\ref{fig: isotropic ellipsoid}d). At the end of discharge ($t=3600$ s, Figs.~\ref{fig: isotropic ellipsoid}e-h), particles under most constraints are nearly fully lithiated, except for the inner-free case. 

These results indicate the \textit{constraint-sensitivity} of isotropic particles: the ellipsoidal morphology alone induces only mild spatial heterogeneity in lithiation (i.e., Fig.~\ref{fig: isotropic ellipsoid}d), but varying mechanical constraints greatly alters the lithiation pathway. Among the various constraints, the inner-free condition most strongly amplifies heterogeneous and anisotropic lithiation.

\begin{figure}[htbp]
    \centering
    \includegraphics[width=0.9\linewidth]{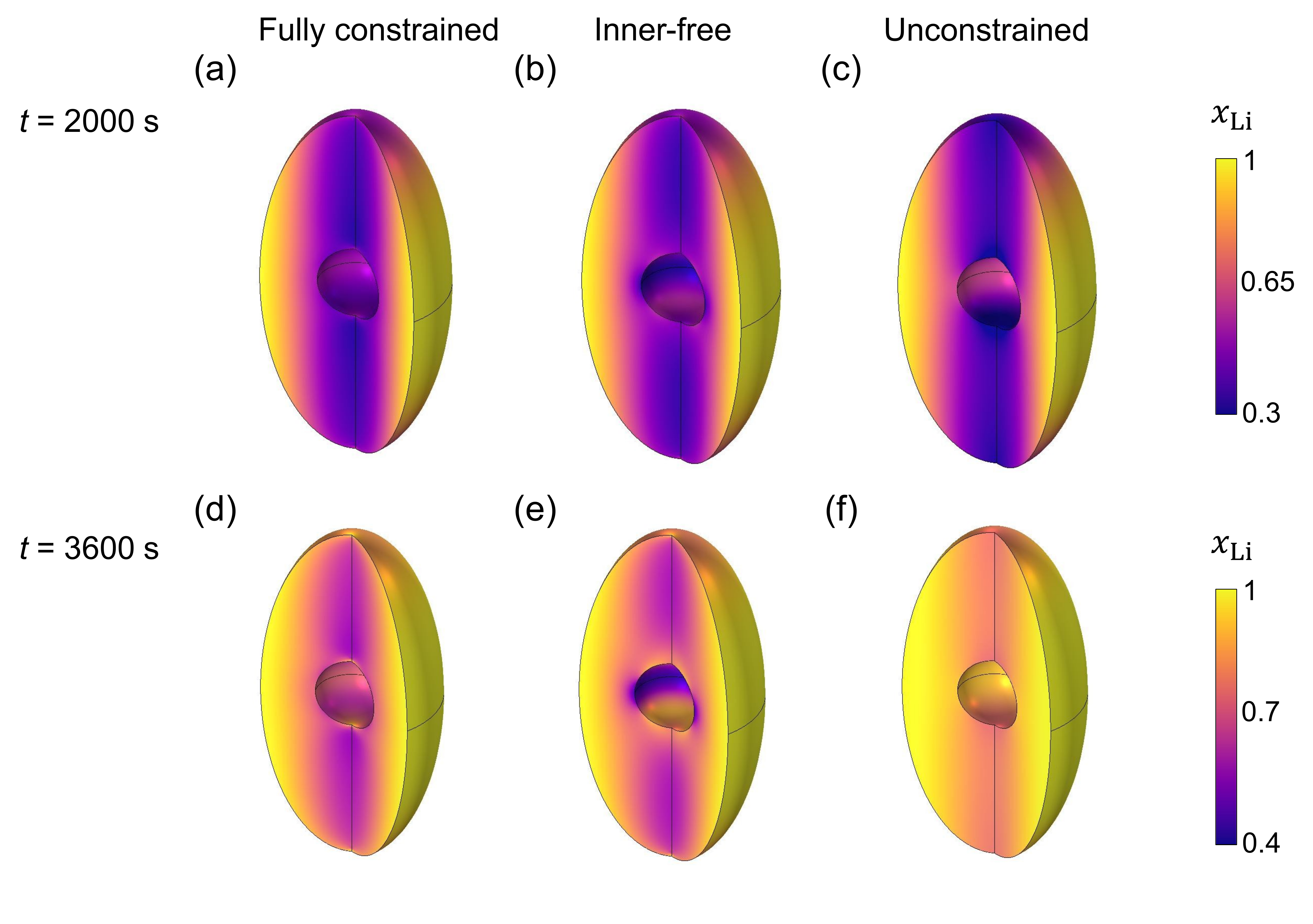}
    \caption{Full fields of normalized Li concentration $x_{\rm Li}=c/c_{\rm max}$ in a transversely isotropic ellipsoid at $t=2000$ s (first row) and $t=3600$ s (second row) under different mechanical constraints: (a)\&(d) fully constrained, (b)\&(e) inner-free, and (c)\&(f) unconstrained.}
    \label{fig: TI ellipsoid}
\end{figure}

Compared to isotropic particles, transversely isotropic particles show much less constraint-sensitivity throughout lithiation (Fig.~\ref{fig: TI ellipsoid}). At $t=2000$ s (Fig.~\ref{fig: TI ellipsoid}a-c), particles under all three different constraints exhibit similar heterogeneous and anisotropic lithiations, with $x_{\rm Li} \approx 1$ near the outer surface along the short axis and $x_{\rm Li} \approx 0.3$ along the long axis. Similar heterogeneous and anisotropic patterns further persist at the end of discharge ($t=3600$ s, Fig.~\ref{fig: TI ellipsoid}d-f), where varying the mechanical constraints only alters the overall magnitude of $x_{\text{Li}}$ but minimally affects the spatial distribution. In other words, unlike in isotropic particles where mechanical constraints exert a strong influence, lithiation in transversely isotropic particles is predominantly governed by the interplay between the ellipsoidal morphology and the transversely isotropic material properties.

To identify the origin of the heterogeneous lithiation pattern in the transversely isotropic particle, we further decompose the anisotropy into transport and chemoelastic contributions. In addition to the fully isotropic and fully transversely isotropic cases, we consider two intermediate material descriptions: (i)\textit{ isotropic diffusivity}, in which the diffusivity tensor is isotropic while the Vegard and elastic stiffness tensors are transversely isotropic, and (ii) \textit{isotropic chemoelasticity}, in which the diffusivity tensor is transversely isotropic while the Vegard and elastic stiffness tensors are both isotropic. The latter groups the Vegard and elastic tensors together as they both arise from the same crystallographic material symmetry and orientation. To quantify the resulting spatial lithiation heterogeneity, we define the normalized spatial standard deviation of Li concentration as
\begin{equation}
    H_c = \frac{\sqrt{\left\langle \left(c-\langle c\rangle\right)^2 \right\rangle}}{c_{\max}-c_0},
\end{equation}
where $\langle \cdot \rangle$ denotes the volume average over the particle domain. 

\begin{figure}
    \centering
    \includegraphics[width=0.9\linewidth]{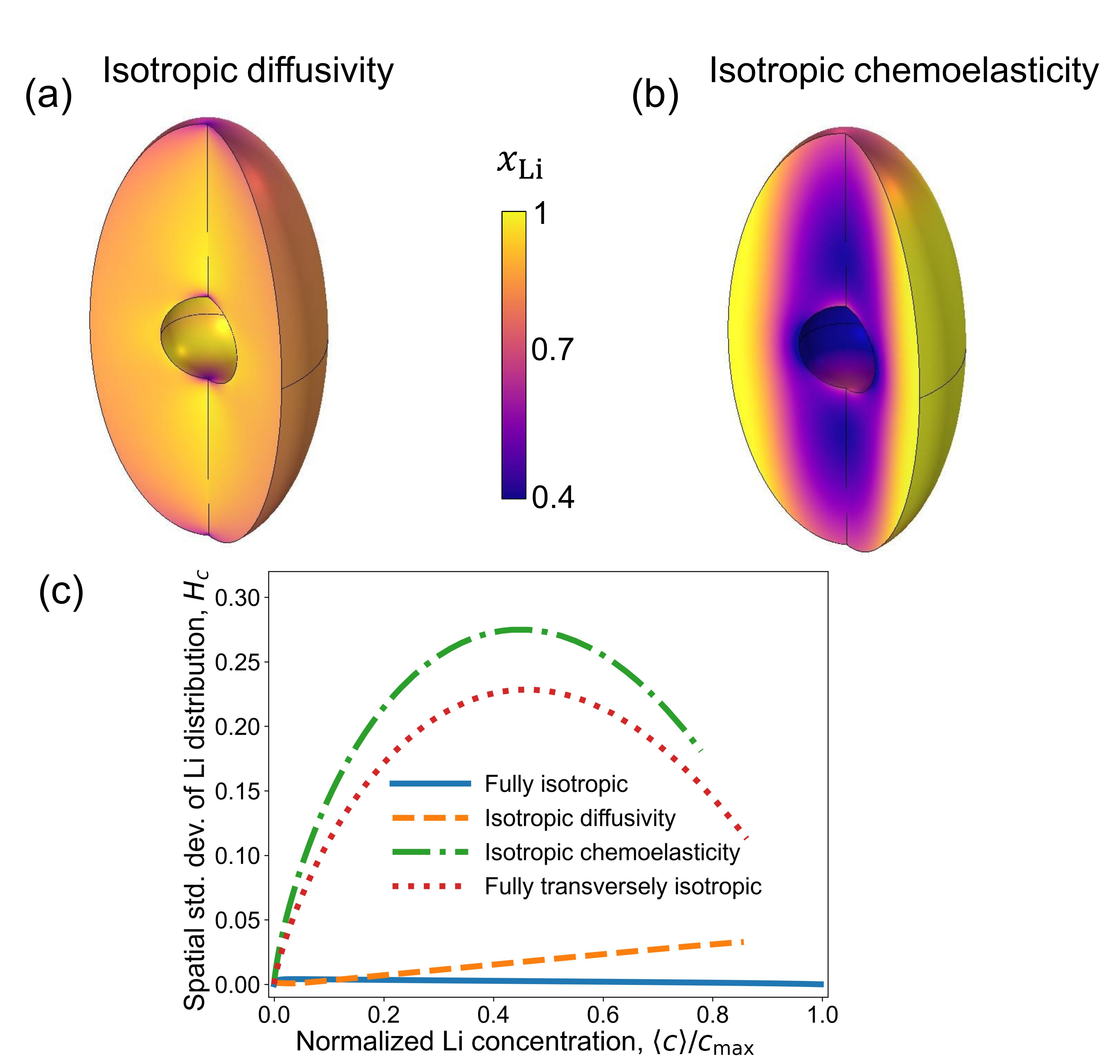}
    \caption{Decomposition of anisotropy into transport and chemoelastic contributions in fully constrained ellipsoidal particles. Representative full fields of normalized Li concentration \(x_{\mathrm{Li}}=c/c_{\max}\) at \(t\) = 2000 s are shown for (a) isotropic diffusivity, in which the diffusivity is isotropic while the Vegard and elastic stiffness tensors are transversely isotropic, and (b) isotropic chemoelasticity, in which the diffusivity tensor is transversely isotropic while the Vegard and elastic stiffness tensors are both isotropic. (c) Evolution of the normalized spatial standard deviation of Li concentration, $H_c$, as a function of normalized volume-averaged Li concentration, \(\langle c\rangle/c_{\max}\), for the fully isotropic, isotropic diffusivity, isotropic chemoelasticity, and fully transversely isotropic cases.}
    \label{fig:isoD_isoEl}
\end{figure}

Figs. \ref{fig:isoD_isoEl}a\&b plot the full fields of normalized Li concentration \(x_{\mathrm{Li}}=c/c_{\max}\) at \(t\) = 2000 s, in the cases of isotropic diffusivity and isotropic chemoelasticity, respectively, under the fully constrained condition. Correspondingly, Fig. \ref{fig:isoD_isoEl}c plots \(H_c\) as a function of the normalized volume-averaged Li concentration, \(\langle c\rangle/c_{\max}\).The fully isotropic case exhibits negligible heterogeneity throughout lithiation, with a maximum value of only \(H_c^{\text{max}}\approx 0.004\). The heterogeneity remains small in the isotropic diffusivity case, reaching \(H_c^{\text{max}}\approx 0.033\). By contrast, the isotropic chemoelasticity case exhibits much stronger heterogeneity with a peak value of \(H_c^{\text{max}}\approx 0.275\), even exceeding that of the fully transversely isotropic case with \(H_c^{\text{max}}\approx 0.228\). This comparison indicates that the pronounced heterogeneous lithiation is governed primarily by the anisotropic diffusivity. The slightly weaker heterogeneity in the fully transversely isotropic case further suggests that the anisotropic chemoelasticity partially counteracts the effect from the anisotropic diffusivity.

\section{Cross-Field Analysis of Chemo-Mechanical Coupling}
\label{sec:cross-field analysis}

To quantify how each mechanism contributes to the resultant transport, we use a projection-based attribution measure. Similar projection concepts have been used in multi-component systems to distinguish contributions that reinforce or oppose a resultant signal \citep{jin_fossil_2026}, while related decomposition approaches have been used in chemical physics to separate resultant forces into physically interpretable contributions \citep{aldossary_force_2023}. 

\subsection{Flux decomposition}
\label{sec:flux decomposition}

To first quantify the contribution of each mechanism to lithiation, we decompose the total Li flux into:
\begin{equation}
\boldsymbol{J}_{\mathrm{tot}}=\boldsymbol{J}_{\mathrm{c}}+\boldsymbol{J}_{\mathrm{m}},
\end{equation}
where 
\begin{equation}\label{eq:J_c}
J_{\mathrm{c},i} = -D_{ij}\frac{c_{\max}}{(c_{\max}-c)}\frac{\partial c}{\partial x_j},
\end{equation}
represents the chemical contribution from the classical Fickian diffusion-reaction,
and
\begin{equation}\label{eq:J_m}
    J_{\mathrm{m},i}=\frac{D_{ij}c}{RT}C_{klmn}\beta_{mn}\frac{\partial}{\partial x_j}(\varepsilon_{kl}-\beta_{kl}\tilde{c}),
\end{equation}
represents the mechanical contribution from the lithiation-induced transformation strain.

To quantify how each mechanism contributes to the resultant transport, we use a projection-based attribution measure. Similar projection concepts have been used in multi-component systems to distinguish contributions that reinforce or oppose a resultant signal \citet{jin_fossil_2026}, while related decomposition approaches have been used in chemical physics to separate resultant forces into physically interpretable contributions \citet{aldossary_force_2023}. We therefore define the \textit{global flux ratio} to project each flux component onto the total Li flux and integrate the contribution over the particle domain:
\begin{equation}
C_{\alpha}(t)=
\frac{
\displaystyle\int_V
\boldsymbol{J}_{\alpha}(\boldsymbol{x},t)\cdot
\boldsymbol{J}_{\mathrm{tot}}(\boldsymbol{x},t)\,\text{d}V
}{
\displaystyle\int_V
\|\boldsymbol{J}_{\mathrm{tot}}(\boldsymbol{x},t)\|^2\,\text{d}V
},
\qquad
\alpha\in\{\mathrm{c},\mathrm{m}\}.
\label{eq:C_global}
\end{equation}
Positive and negative values of $C_{\alpha}$ indicate global reinforcement and reduction of the total flux, respectively, by the corresponding flux component. Because $C_{\mathrm{c}}+C_{\mathrm{m}}=1$, values greater than unity or negative indicate that individual components are partially canceled by other components.

\begin{figure}[htbp]
    \centering
    \includegraphics[width=0.9\linewidth]{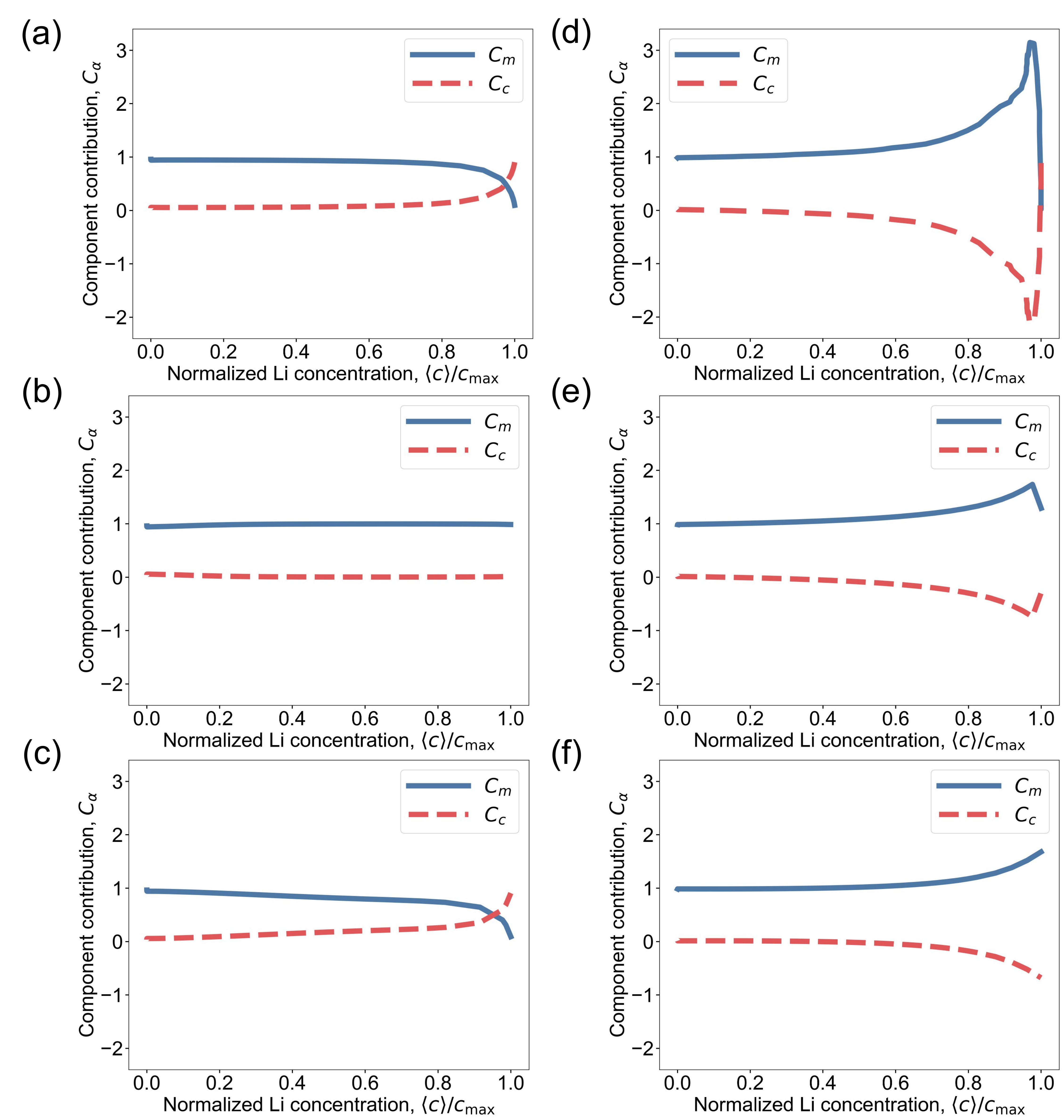}
    \caption{Evolution of global flux ratio \(C_\alpha\) as a function of normalized volume-averaged Li concentration, $\langle c \rangle /c_{\max}$, in (a) fully constrained, (b) inner-free, and (c) unconstrained isotropic ellipsoidal particles, as well as (d) fully constrained, (e) inner-free, and (f) unconstrained transversely isotropic ellipsoidal particles.}
    \label{fig:C_alpha}
\end{figure}

Fig.~\ref{fig:C_alpha} plots the evolution of $C_{\mathrm{c}}$ and $C_{\mathrm{m}}$ in ellipsoidal particles with different material properties and mechanical constraints. In all cases, the mechanical contribution (from lithiation-induced transformation strain) dominates over the chemical contribution (from Fickian diffusion-reaction). This dominance reverses near the end of lithiation $\langle c \rangle /c_{\max}=1$ in some cases (Figs. \ref{fig:C_alpha}a, c, and d, corresponding to fully constrained and unconstrained isotropic particles and a fully constrained transversely isotropic particle), but the total flux itself is small by that point. The observation, that the mechanical contribution dominates across all constraints and material properties, suggests that this behavior is an intrinsic characteristic of the ellipsoidal morphology.

To further establish the morphological origin of this difference, we systematically vary the particle aspect ratio from spherical to elongated ellipsoidal while maintaining the same isotropic material properties, inner-free mechanical constraint, particle volume, and cavity geometry, as detailed in Appendix B. As the aspect ratio decreases, the short- and long-axis concentration profiles progressively diverge, accompanied by a systematic increase in the signed mechanical contribution to Li transport (Fig.~\ref{app fig:aspect ratio}). These results establish a continuous transition between the spherical and ellipsoidal cases and demonstrate that increasing geometric anisotropy amplifies both directional lithiation heterogeneity and the relative mechanical contribution to Li transport.

\subsection{Correlation between concentration and deformation}
\label{sec:spatial correlation}

We next investigate the spatial correspondence between chemical (Li concentration) and mechanical (strain) fields to assess when deformation may retain useful information about the underlying lithiation state, or vice versa. Simultaneous measurements of Li concentration and strain have demonstrated spatial chemo-mechanical correspondence during intercalation \citep{yang_-situ_2019}, while cell-level stress and strain signals have been used to estimate electrochemical states such as the state of charge (SOC) and state of health (SOH) \citep{cannarella_state_2014, dai_state_2017, ganguli_embedded_2017}. More broadly, full-field mechanical measurements are widely used to infer otherwise inaccessible material information \citep{avril_overview_2008}, and recent work suggests that strain fields can encode hidden reaction-diffusion kinetics and thermodynamic quantities in coupled chemo-mechanical systems \citep{ihuaenyi_learning_2025}. Motivated by these observations, we use the Pearson correlation coefficient as a forward measure of field-to-field correspondence. 

To do so, we introduce the Pearson correlation metric to quantify spatial co-variation between the volumetric strain field $\varepsilon_v(\boldsymbol{x},t)=\varepsilon_{rr}+\varepsilon_{\theta\theta}+\varepsilon_{zz}$ and the Li concentration field $c(\boldsymbol{x},t)$ \citep{lee_rodgers_thirteen_1988}:
\begin{equation}
\rho_{\varepsilon c}(t)=
\frac{
\displaystyle \int_{V}
\bigl(\varepsilon_v(\boldsymbol{x},t)-\bar{\varepsilon}_v(t)\bigr)
\bigl(c(\boldsymbol{x},t)-\bar{c}(t)\bigr)\,\text{d}V
}{
\displaystyle
\sqrt{
\int_{V}\bigl(\varepsilon_v(\boldsymbol{x},t)-\bar{\varepsilon}_v(t)\bigr)^{2}\,\text{d}V\;
\int_{V}\bigl(c(\boldsymbol{x},t)-\bar{c}(t)\bigr)^{2}\,\text{d}V
}
},
\label{eq: pearson corr}
\end{equation}
where the overbar denotes the spatial average over the particle volume.

\begin{figure}
    \centering
    \includegraphics[width=0.9\linewidth]{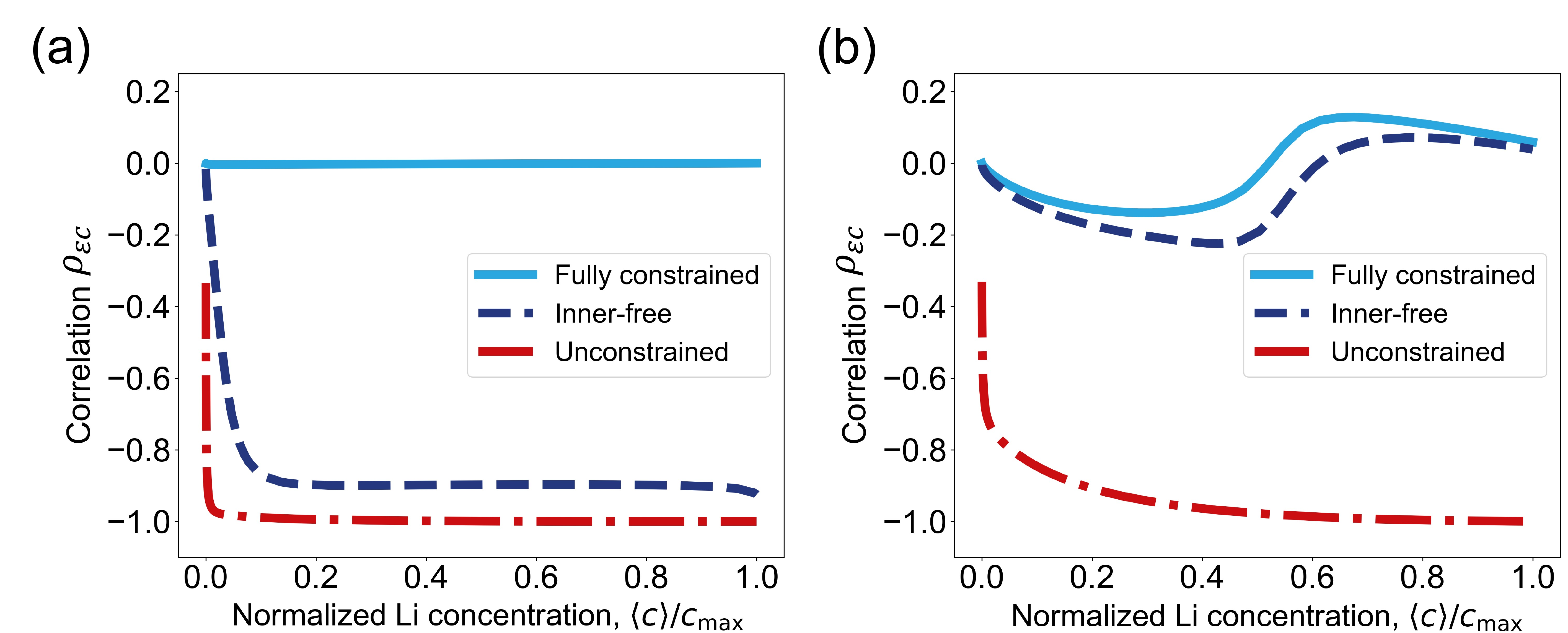}
    \caption{Pearson correlation metric to quantify spatial co-variation between the volumetric strain field $\varepsilon_v(\boldsymbol{x},t)=\varepsilon_{rr}+\varepsilon_{\theta\theta}+\varepsilon_{zz}$ and the Li concentration field $c(\boldsymbol{x},t)$, for (a) isotropic and (b) transversely isotropic ellipsoidal particles under different mechanical constraints.}
    \label{fig: corr between c & strain}
\end{figure}

We compute $\rho_{\varepsilon c}(t)$ for isotropic and transversely isotropic ellipsoidal particles (Figs~\ref{fig: corr between c & strain}a\&b, respectively) under different mechanical constraints. Overall, $\rho_{\varepsilon c}(t)$ is strongest in the unconstrained case, intermediate in the inner-free case, and weakest in the fully constrained case. Indeed, in unconstrained particles, $\varepsilon_v(\boldsymbol{x},t)$ is dominated by the \textit{local} lithiation-induced transformation strain, leading to near-fully anti-correlated two fields with $\rho_{\varepsilon c}\approx-1$ throughout lithiation, consistent with \citep{ihuaenyi_learning_2025}. This anti-correlation reflects the negative Vegard response: higher Li concentration produces more negative volumetric strain (see values of $\beta$ in Table \ref{table: modeling parameters}), corresponding to greater volumetric shrinkage. In fully constrained particles, by contrast, $\varepsilon_v(\boldsymbol{x},t)$ is governed by the \textit{nonlocal} elastic response to the transformation-strain field under the stronger constraints. Material anisotropy further modulates the correlation, as the anisotropic stiffness, Vegard, and diffusivity tensors introduce additional degrees of freedom that reshape both the concentration field and the elastic response.

\begin{figure}
    \centering
    \includegraphics[width=1\linewidth]{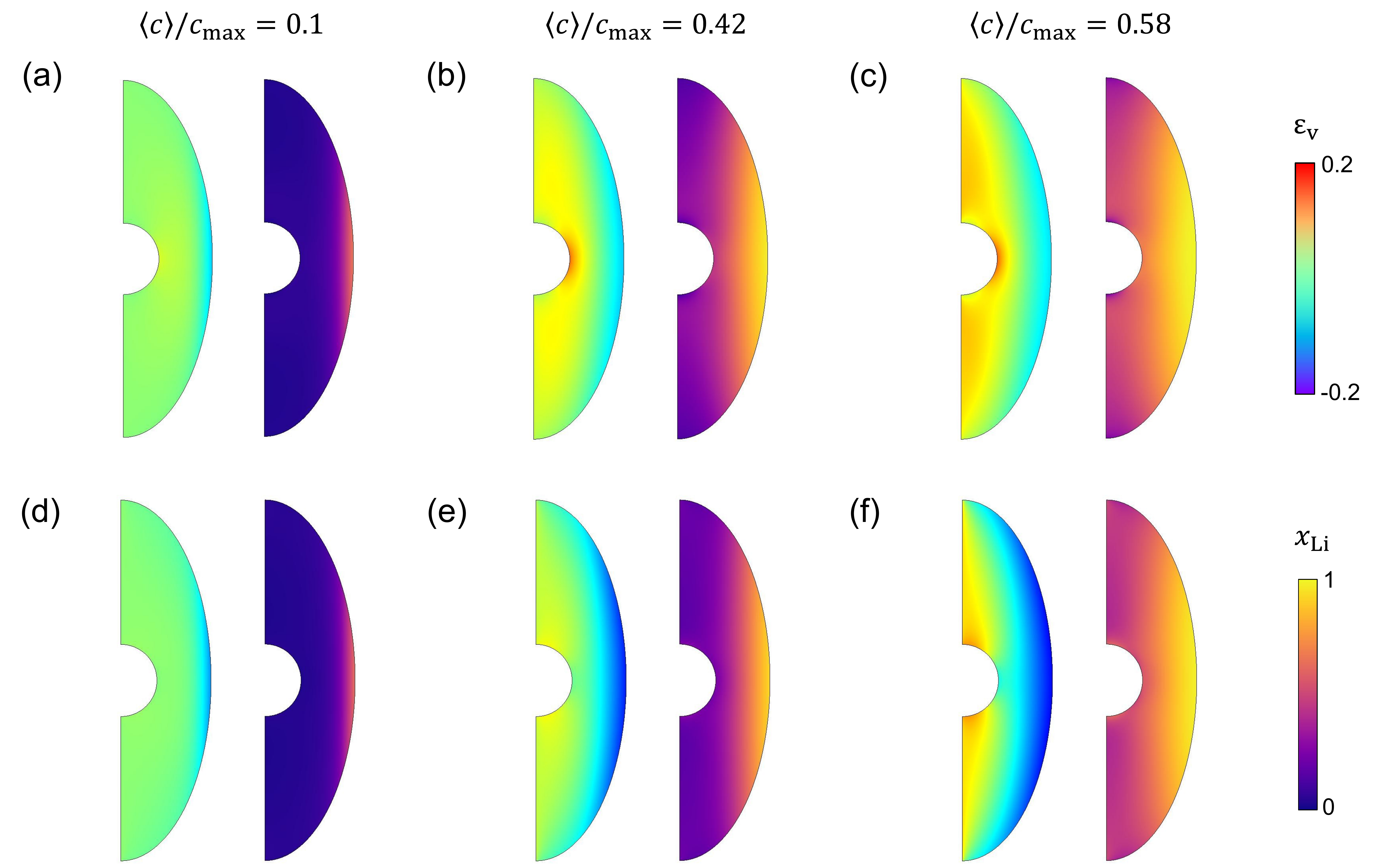}
    \caption{Representative fields of volumetric strain $\varepsilon_v(\boldsymbol{x},t)$ (left) and normalized Li concentration $x_{\mathrm{Li}}=c/c_{\max}$ (right) during lithiation, $\langle c\rangle/c_{\max}=$0.10, 0.42, and 0.58, in transversely isotropic ellipsoidal particles. (a-c) show fully constrained particles. (d–f) show unconstrained particles.}
    \label{fig:corr snapshots}
\end{figure}

These physical pictures are confirmed by the corresponding strain and concentration fields during lithiation (left and right of Figs.~\ref{fig:corr snapshots}a-f). Fully constrained transversely isotropic particles exhibit no clear visual correlation between the two fields throughout lithiation (Figs.~\ref{fig:corr snapshots}a-c): the concentration field (right) develops an overall horizontal gradient, whereas the strain field (left) is more multi-axially distributed owing to the nonlocal elastic response. As a result, regions of high concentration do not coincide with regions of high strain. Unconstrained particles, by contrast, show a clear visual correlation (Figs.~\ref{fig:corr snapshots}d-f): the strain field closely tracks the concentration field throughout lithiation, with higher concentrations yielding more negative strains.

These results show that the concentration-deformation correspondence is condition-dependent rather than universal. Under weak mechanical constraint, deformation largely follows the local transformation strain and can therefore retain strong spatial information about lithiation. In contrast, under strong constraint, the nonlocal elastic response substantially weakens this correspondence. The analysis thus identifies conditions under which mechanical and electrochemical fields may serve as useful proxies for each other.

\section{Morphology Optimization} \label{sec:opt}

The preceding sections show that particle morphologies, material properties, and mechanical constraints affect chemo-mechanical behavior through distinct but coupled mechanisms. The analysis suggests that the chemo-mechanical performance of a particle cannot be optimized against a single objective or by tuning a single material or geometric parameter. Therefore, we formulate a multi-objective optimization problem with a set of design variables that describe the particle morphology including its geometry and crystal orientation. As a demonstration, we take the inner-free transversely isotropic ellipsoidal particle under half-cycle lithiation ($t=3600$ s) and apply the proposed framework. The optimization seeks to maximize lithiation capacity while suppressing the lithiation-induced peak tensile stress.

\subsection{Formulation of the optimization problem}

We start with defining a vector of optimization parameters, $\boldsymbol{\theta}=[a_{\text{in}}, c_{\text{in}},p_y,\alpha]^{\text{T}}$, where $a_{\text{in}}$ and $c_{\text{in}}$ are the semi-axes of the internal ellipsoidal void of the hollow ellipsoidal particle, $p_y$ describes the centroid position of the void along the $z$-axis (Figure~\ref{fig: energy comparison}d left), and $\alpha$ represents the angle between the crystallographic $c$-axis and the short axis of the ellipsoid. Throughout the optimization, we restrict $\alpha$ within the first quadrant $0 \leq \alpha \leq \pi/2$ due to symmetry, and fix the outer semi-axes of the ellipsoid as $a_{\mathrm{out}} = 5\,\mu\mathrm{m}$ and $c_{\mathrm{out}} = 10\,\mu\mathrm{m}$.

We take the radially aligned crystal configuration in the transversely isotropic ellipsoidal particle as the baseline (Section.~\ref{sec:ellip}). For an arbitrarily prescribed $\alpha$, the diffusivity, Vegard, and stiffness tensors are expressed as
\begin{equation}
    \boldsymbol{D} = \boldsymbol{R}(\alpha) \, \boldsymbol{D}^{(\text{base})} \, \boldsymbol{R}^{T}(\alpha),
\end{equation}
\begin{equation}
    \boldsymbol{\beta} = \boldsymbol{R}(\alpha) \,\boldsymbol{\beta}^{(\text{base})} \, \boldsymbol{R}^{T}(\alpha),
\end{equation}
and
\begin{equation}
\boldsymbol{C} = \boldsymbol{T}(\alpha) \, \boldsymbol{C}^{(\text{base})} \, \boldsymbol{T}^{T}(\alpha),
\end{equation}
where $\boldsymbol{D}^{(\text{base})}$, $\boldsymbol{\beta}^{(\text{base})}$, and $\boldsymbol{C}^{(\text{base})}$ are the corresponding quantities in the baseline particle, and the rotation matrices are
\begin{equation}
\boldsymbol{R}(\alpha) =
\begin{bmatrix}
\cos\alpha & 0 & \sin\alpha \\
0 & 1 & 0 \\
-\sin\alpha & 0 & \cos\alpha
\end{bmatrix},
\end{equation}
and
\begin{equation*}
\boldsymbol{T}(\alpha) =
\begin{bmatrix}
\cos^2\alpha & \sin^2\alpha & 2\cos\alpha\sin\alpha \\
\sin^2\alpha & \cos^2\alpha & -2\cos\alpha\sin\alpha \\
-\cos\alpha\sin\alpha & \cos\alpha\sin\alpha & \cos^2\alpha-\sin^2\alpha
\end{bmatrix}.
\end{equation*}
The flux in a particle with a prescribed $\alpha$ is given by Eq.~\ref{eq:oriented flux} in Appendix \ref{app:analytical for ellips}.

The morphology optimization is formulated as a constrained multi-objective problem that seeks the optimal parameter vector $\boldsymbol{\theta}^*$ following:
\begin{align}
\boldsymbol{\theta}^* &= \arg\min_{\boldsymbol{\theta} \in \mathcal{D}} \left\{  -\bar{c}_{\text{ave}}(\boldsymbol{\theta}),\,  \,\sigma_{1,\max}(\boldsymbol{\theta}) \right\}, \label{eq:optimization_problem}
\end{align}
subject to geometric admissibility that ensures ellipsoidal morphology:
\begin{align}
&a_{\text{in}} > 0, \quad c_{\text{in}} > 0, \label{eq:physical_bounds} \\
&p_x^2/a_{\text{out}}^2 + p_y^2/c_{\text{out}}^2 < 1, \label{eq:centroid_enclosure} \\
&|p_x| + a_{\text{in}} \leq a_{\text{out}} - t_{\min}, \quad |p_y| + c_{\text{in}} \leq c_{\text{out}} - t_{\min}, \label{eq:wall_thickness} \\
&\frac{(p_x + a_{\text{in}} \cos\theta)^2}{a_{\text{out}}^2} + \frac{(p_y + c_{\text{in}} \sin\theta)^2}{c_{\text{out}}^2} < 1 - \frac{t_{\min}}{L_c} \quad \forall \theta \in [0, 2\pi], \label{eq:void_enclosure} \\
&V_{\mathrm{in}} = \frac{4\pi}{3}a_{\mathrm{in}}^2c_{\mathrm{in}} = 0.032V_{\mathrm{out}},
\label{eq:volume_bounds} 
\end{align}
where $\bar{c}_{\text{ave}}=\langle c \rangle /c_{\max}$ is the normalized average lithium concentration after half-cycle lithiation and $\sigma_{1,\max}$ is the peak maximum principal stress attained over the entire lithiation history, normalized by the elastic modulus $E=123.6$ GPa. $\mathcal{D} = [\lambda_{\min}, \lambda_{\max}]^{N}$ denotes the admissible design space with component-wise lower ($\lambda_{\min}$) and upper bounds ($\lambda_{\max}$) on each of the $N = 4$ design variables. Eq.~\ref{eq:physical_bounds} requires strictly positive void semi-axes, preventing degenerate zero-volume voids. Eq.~\ref{eq:centroid_enclosure} ensures that the void is inside the particle, where $p_x = 0$ represents an axisymmetric void about the $z$-axis. Eq.~\ref{eq:wall_thickness} enforces a minimum wall thickness $t_{\min}$ of the hollow particle along each principal axis, which is a necessary but not sufficient condition for full enclosure. Eq.~\ref{eq:void_enclosure} requires that every point on the parametric void boundary satisfies the outer ellipsoid inequality with a margin $t_{\min}/L_c$, where $L_c = \sqrt{a_{\mathrm{out}}^2 + c_{\mathrm{out}}^2}$ is the characteristic length of the outer ellipsoid and $t_{\min}=0.1\ \mu\text{m}$. This inequality ensures a minimum clearance between the void surface and the outer boundary of the ellipsoid. Eq.~\ref{eq:volume_bounds} prescribes a constant void volume throughout the optimization, such that the active-material volume, \(V_{\mathrm{act}}=V_{\mathrm{out}}-V_{\mathrm{in}}\) is identical for all candidate morphologies. Consequently, the total amount of intercalated Li is proportional to \({\bar{c}}_{\mathrm{ave}}\), and maximizing it is equivalent to maximizing the Li inventory at the prescribed lithiation time.

The multi-objective functional in Eq.~\ref{eq:optimization_problem} aims to maximize the final normalized average Li concentration   $\bar{c}_{\text{ave}}$ while minimizing the peak maximum principal stress $\sigma_{1,\max}$ encountered throughout lithiation. The former represents the electrochemical capacity and state of charge, and the latter indicates the most severe tensile-stress state experienced by the particle during the transient process and therefore serves as a mechanical degradation indicator. The optimization thus promotes lithium uptake while preserving structural integrity. As identified in the preceding sections, we expect a design trade-off between the two objectives: greater Li uptake increases the lithiation-induced transformation strain and hence the resulting stress and strain fields. As a result, the optimization yields multiple possible results that form a Pareto front.

\subsection{Optimization Procedure}
We solve the optimization problem using sequential model-based optimization (SMBO) \citep{hutter2011sequential}, wherein a probabilistic surrogate iteratively approximates the multi-objective functional Eq. \ref{eq:optimization_problem} to direct sampling toward regions of high potential. This approach also reduces computational cost by minimizing the number of expensive function evaluations required for convergence.

We employ the Tree-structured Parzen Estimator (TPE)~\citep{bergstra2011algorithms} as a surrogate within the SMBO framework. TPE reformulates the conventional expected improvement (EI) acquisition function via Bayes’ rule. Rather than modeling the objective function directly, TPE partitions the observed data at a threshold into superior and inferior configurations. Separate probability density functions are fitted to each partition using adaptive Parzen window estimators. This recasts the EI criterion as a likelihood-based sampling problem, biasing new candidates toward regions with higher probability of improvement relative to current best evaluations. Mathematical details of TPE are provided in~\citep{bergstra2011algorithms}, with extensions to information-driven mechanical specimen design described in~\citep{ihuaenyi2024seeking, ihuaenyi2025mechanics, ihuaenyi2026mechanics}. Implementation proceeds via the Optuna framework~\citep{akiba2019optuna}, which provides an efficient realization of TPE, including multivariate kernel density estimation and robust acquisition optimization. The study employs a computational budget of 350 trials. In each trial, a candidate parameter vector $\boldsymbol{\theta}$ is sampled from uniform priors, used to generate the particle geometry, and executed in the finite element solver. The final volume-averaged Li concentration is used to evaluate \(\bar c_{\mathrm{ave}}\), while the spatial maximum of \(\sigma_1\) is evaluated throughout the time history and its largest value is retained as \(\sigma_{1,\max}\). The algorithm maintains a complete record of evaluated configurations and objective values, enabling post-optimization analysis and identification of the optimal morphology.

\subsection{Optimization results}

\begin{figure}[h!]
    \centering
    \includegraphics[width=1.0\linewidth]{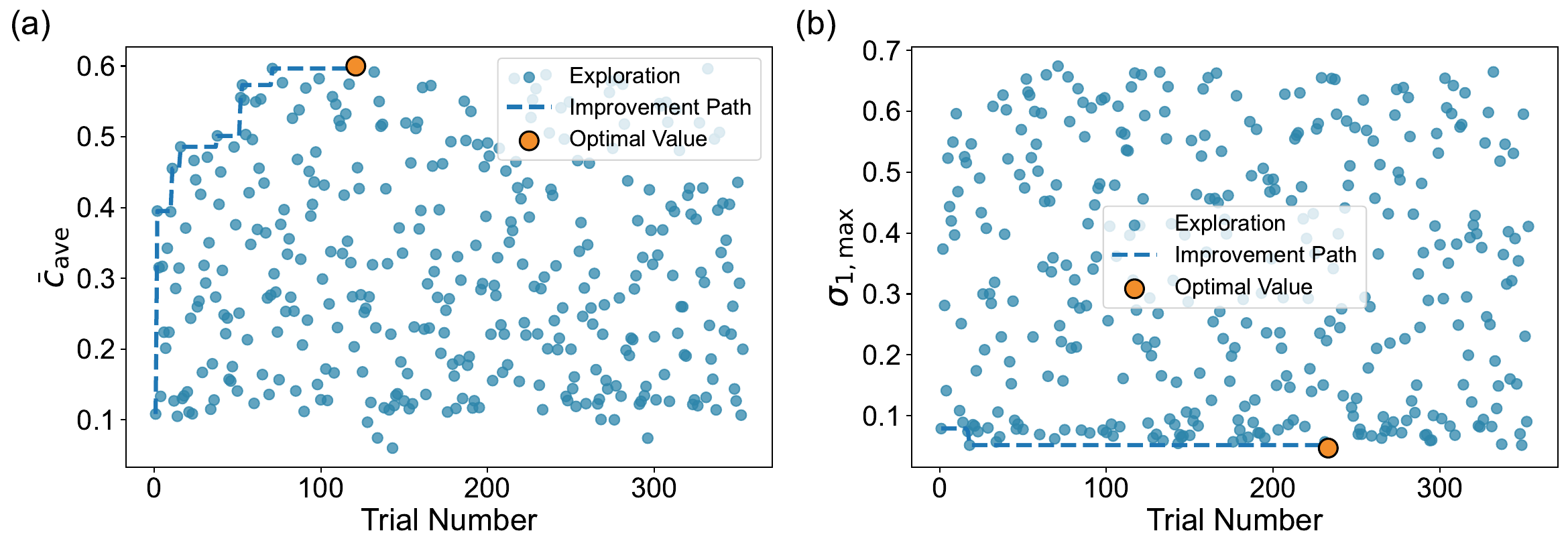}
    \caption{Sequential trial histories of morphology optimization of the inner-free transversely isotropic ellipsoidal particle under half-cycle lithiation ($t=3600$ s). (a) Maximizing the normalized average lithium concentration $\bar{c}_{\text{ave}}$. (b) Minimizing the normalized maximum principal stress $\sigma_{1,\max}$. Blue dots denote explored designs, dashed lines represent improvement paths, and yellow dots denote selected optimal values. Each optimization procedure includes 350 evaluations.}
    \label{fig: opt}
\end{figure}

Fig.~\ref{fig: opt} shows the sequential trial histories produced by the Bayesian optimization procedure over 350 evaluations, each maximizing $\bar{c}_{\text{ave}}$ or minimizing $\sigma_{1,\max}$. Each exploration point corresponds to a distinct candidate morphology sampled from the admissible design space $\mathcal{D}$. The dashed lines denote the best-so-far value for each objective and visualize convergence trends. The optimal $\bar{c}_{\text{ave}}=0.60$ is reached relatively early and remains close to the upper envelope of the sampled designs, indicating that high-lithiation morphologies are accessible within $\mathcal{D}$. In contrast, $\sigma_{1,\max}$ continues to improve over a broader range of trials before reaching its minimum $\sigma_{1,\max}=0.047$, suggesting that stress mitigation is more sensitive to the specific combination of optimization parameters. 


\begin{table}[h!]
\caption{Parameters of optimized solutions}
\label{tab:opt param}
\centering
\begin{tabular}{lcccc}
\toprule
\text{Optimized solution} & $\boldsymbol{p_y}$ & $\boldsymbol{a_{\text{in}}}$ & $\boldsymbol{c_{\text{in}}}$ & $\boldsymbol{\alpha}$ \\
\midrule
Best compromise & $-1.68$ & $1.03$ & $7.54$ & $0.3^\circ$ \\
Highest $\bar{c}_{\text{ave}}$& $0.59$ & $3.19$ & $0.79$ & $27.15^\circ$ \\
Lowest $\sigma_{1,\max}$& $-1.13$ & $1.71$ & $2.72$ & $0.67^\circ$ \\
\bottomrule
\end{tabular}
\end{table}

\begin{figure}
    \centering
    \includegraphics[width=1\linewidth]{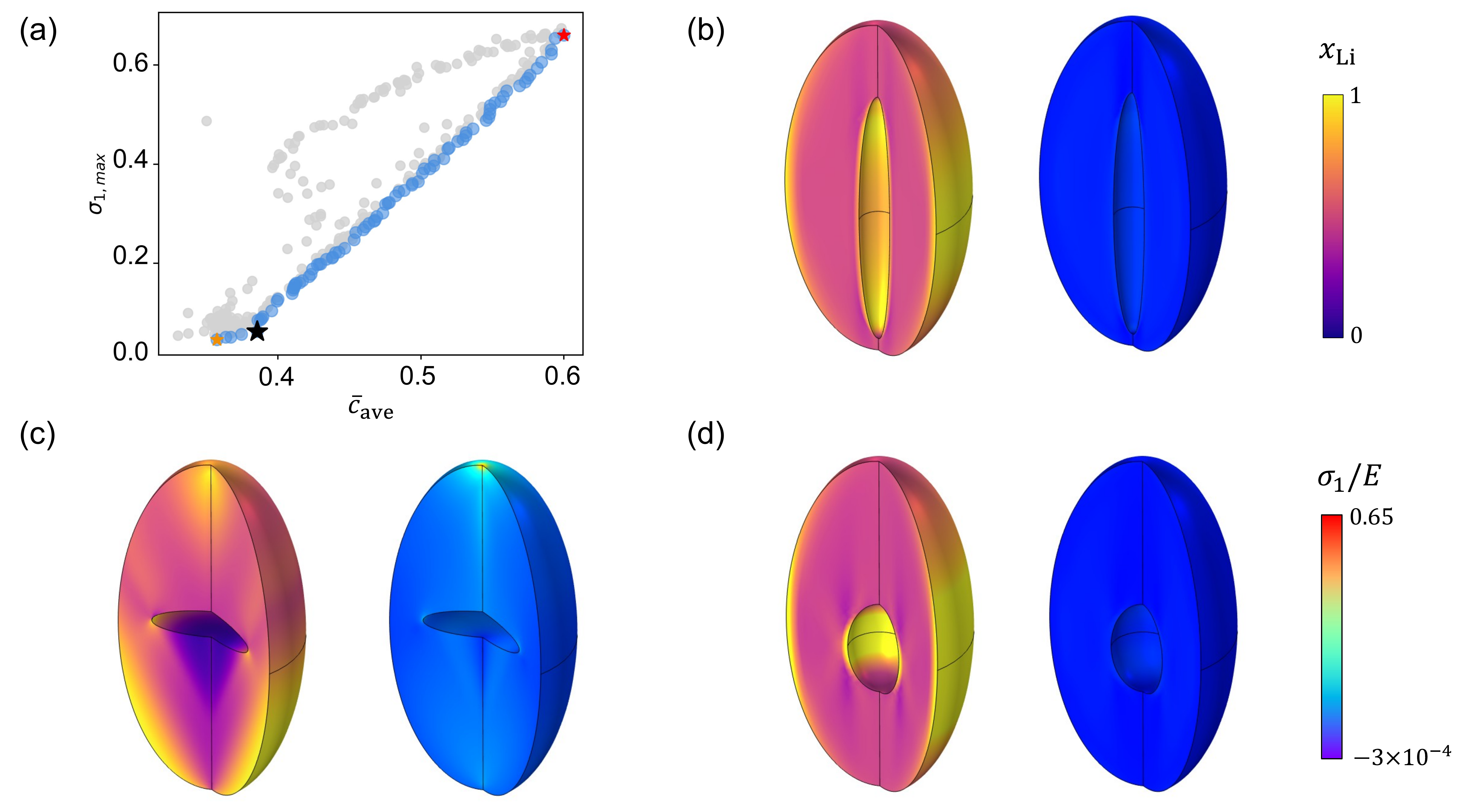}
    \caption{Multi-objective optimization results from Pareto analysis and the corresponding morphologies with field distributions. (a) Trade-off between the final average Li concentration $\bar{c}_{\text{ave}}$, and the peak maximum principal stress $\sigma_{1,\max}$ attained over the entire lithiation history. Gray dots denote all explored designs. Blue dots represent the family of non-dominated morphologies from Pareto analysis. The black, red, and orange stars represent (b) the best-compromise morphology, (c) the morphology with the highest $\bar{c}_{\text{ave}}$, and (d) the morphology with the lowest $\sigma_{1,\max}$, respectively. (b)-(d) plot their corresponding fields of normalized Li concentration $x_{\mathrm{Li}}=c/c_{\max}$ (left) and normalized maximum principal stress $\sigma_1$ (right).}
    \label{fig:opt fields}
\end{figure}

While the sequential search converges to morphologies that individually maximize $\bar{c}_{\text{ave}}$ or minimize $\sigma_{1,\max}$, the two optima conflict and cannot coexist in a single morphology. As shown in Fig. \ref{fig:opt fields}a, larger $\bar{c}_{\text{ave}}$ generally accompanies larger $\sigma_{1,\max}$, establishing an intrinsic trade-off. Such a trade-off is a defining characteristic of multi-objective optimization problems, in which no single design simultaneously optimizes all objectives, requiring a generalized solution concept. 

One appropriate framework that addresses this conflict is Pareto optimality \citep{deb2001multi, deb2002fast}, where a design $\boldsymbol{\theta}^*$ is Pareto optimal if no alternative feasible design improves at least one objective without degrading another. The set of all such non-dominated solutions constitutes a Pareto front, which characterizes the complete envelope of achievable trade-offs in the objective space. Non-dominated solutions are identified via exhaustive pairwise dominance checking. A candidate $\boldsymbol{\theta}_i$ is eliminated if any $\boldsymbol{\theta}_j$ satisfies:
\begin{equation}
    f_k(\boldsymbol{\theta}_j) \leq f_k(\boldsymbol{\theta}_i) \quad \forall\, k, \qquad \text{and} \qquad f_k(\boldsymbol{\theta}_j) < f_k(\boldsymbol{\theta}_i) \quad \text{for at least one } k,
\label{eqn: Pareto analysis}
\end{equation}
where $ f_k$ represents the $k$-th objective (here $k=$ 1 or 2) and maximization objectives are converted to equivalent minimization form by sign negation. Among the resulting Pareto-optimal set, a best-compromise design is selected by maximizing a balanced utility score $\mathcal{U}$, computed as the arithmetic mean of per-objective normalized sub-scores \citep{deb2001multi}:
\begin{equation}
    \mathcal{U}(\boldsymbol{\theta}_i) = \frac{1}{K} \sum_{k=1}^{K} s_k(\boldsymbol{\theta}_i), \qquad
    s_k(\boldsymbol{\theta}_i) = \frac{f_k(\boldsymbol{\theta}_i) - f_k^{\mathrm{worst}}}{f_k^{\mathrm{best}} - f_k^{\mathrm{worst}}},
\label{eqn: best-compromise Pareto}
\end{equation}
where $K = 2$ is the number of objectives in the current problem, $f_k^{\mathrm{best}}$ and $f_k^{\mathrm{worst}}$ denote the best and worst observed values of objective $k$ across all trials, respectively, and $s_k$ is oriented so that higher values are preferred for all objectives. 

Following Eqs. \ref{eqn: Pareto analysis}\&\ref{eqn: best-compromise Pareto}, we perform Pareto analysis to identify the family of non-dominated morphologies (blue dots in Fig. \ref{fig:opt fields}a) and extract the best-compromise morphology from them (Fig. \ref{fig:opt fields}a black star). The best-compromise morphology achieves $\bar{c}_{\text{ave}} = 0.4$ and $\sigma_{1,\max} = 0.063$,  neither of which is individually optimal. For comparison, the morphologies with the highest $\bar{c}_{\text{ave}} =0.6$ (with $\sigma_{1,\max} = 0.63$) and the lowest $\sigma_{1,\max} = 0.046$ (with $\bar{c}_{\text{ave}} =0.36$) are also highlighted (red and orange stars in Fig. \ref{fig:opt fields}a, respectively). An important feature of the mechanical objective is that the peak principal stress does not simply correspond to the final lithiation state. For the optimized morphologies, the largest values of the spatial maximum principal stress generally occur at intermediate lithiation states around \(\bar{c}_{\mathrm{ave}}\approx0.5\). This range also corresponds to the stage at which the Li concentration distribution exhibits the strongest spatial heterogeneity. The coincidence of these two features indicates that the critical tensile stress is closely associated with transient concentration gradients and the resulting heterogeneous transformation strain. As lithiation proceeds beyond this state, the concentration distribution becomes more homogeneous, reducing the degree of local transformation-strain mismatch even though the average Li concentration continues to increase. Consequently, evaluating the stress objective over the full lithiation history is necessary to capture the mechanically most critical state. The design parameters of these three optimized solutions are listed in Table \ref{tab:opt param}.

We also plot the fields of normalized Li concentration $x_{\mathrm{Li}}=c/c_{\max}$ (left) and normalized maximum principal stress $\sigma_1$ (right) for the optimized morphologies corresponding to the best-compromise design (Fig. \ref{fig:opt fields}b), the highest $\bar{c}_{\text{ave}}$ (Fig. \ref{fig:opt fields}c), and the lowest $\sigma_{1,\max}$ (Fig. \ref{fig:opt fields}d). These fields are consistent with the optimized $\bar{c}_{\text{ave}} $ and $\sigma_{1,\max} $ values in each case. The results further illustrate the non-intuitive nature of the optimal particle morphology, which is governed jointly by geometric parameters and crystal orientation.

The optimization above explores a theoretical design space in which internal cavity geometry and crystallographic orientation can be varied independently. Although arbitrary combinations of these variables may not presently be manufactured with precise control, experimental studies on LCO demonstrate that each of the principal design features considered here can be influenced through synthesis. The external LCO particle morphology can be tuned through precursor design and hydrothermal processing, with spherical, plate-like, and other anisotropic morphologies experimentally demonstrated  \citep{ying_preparation_2004, hamao_effect_2024}. The preferential crystallographic orientation can also be introduced through facet-controlled growth and directional assembly, including preferentially exposed \(\{010\}\) facets and aligned nanoplatelet architectures  \citep{wu_flakelike_2016, zhou_hierarchical_2020}. In addition, hollow LCO particles containing internal cavities have been produced by spray pyrolysis, demonstrating that internal free volume can be introduced and tuned through the precursor chemistry and processing conditions \citep{almazrouei_synthesis_2024}.

These experimental demonstrations support the physical relevance of the individual design variables considered here, but they do not imply that the exact optimized combination of cavity shape, cavity position, and crystal orientation can currently be fabricated independently. The Pareto-optimal morphologies should therefore be interpreted as mechanics-informed target architectures, which identify favorable geometric and crystallographic trends for reducing stress while maintaining lithium uptake. Experimentally realizable particles are expected to occupy a synthesis-constrained subset of the theoretical design space considered here. Therefore, incorporating synthesis-dependent constraints into the optimization constitutes an important direction for future work.

\section{Conclusion}
\label{sec:conclusion}

We have shown that the chemo-mechanical response of an ion-intercalation active particle during lithiation is governed not by particle geometry, material property, or mechanical constraint alone, but by their coupled interaction. In spherical particles with isotropic properties, the mechanical contribution dominates over much of lithiation but becomes secondary to the chemical contribution near full lithiation. In cylindrical particles, axial constraint substantially raises the induced elastic energy, while the opposing signs of the basal-plane and c-axis Vegard coefficients in transversely isotropic materials effectively ameliorate the elastic energy by partially accommodating the lithiation-induced deformation. In ellipsoidal particles, isotropic materials are constraint-sensitive: varying the level of mechanical constraints produces markedly different Li concentration distributions. Transversely isotropic ellipsoidal particles, by contrast, exhibit persistent heterogeneous lithiation governed by anisotropic diffusivity, with the mechanical constraints modifying the magnitude but not the spatial pattern of Li distribution. Cross-field analysis reinforces these observations: flux decomposition shows that the mechanical contribution to Li transport is significant in both spherical and ellipsoidal particles, while increasing geometric anisotropy generally enhances the relative mechanical contribution and directional lithiation heterogeneity. The Pearson correlation between Li concentration and volumetric strain transitions from strong anti-correlation in unconstrained particles to weak correlation under full constraints. Bayesian optimization of hollow ellipsoidal particles further demonstrates how these mechanistic insights translate into morphology design. The optimization identifies a Pareto front of non-dominated morphologies that trade off average Li concentration against maximum principal stress, from which a best-compromise design can be extracted by simultaneously tuning void geometry and crystal orientation, representing mechanics-informed target architectures.

These conclusions are subject to the assumptions of single-phase intercalation, ideal Li-vacancy mixing, small-strain elasticity, and concentration-independent material properties. In addition, the present framework does not explicitly resolve primary grain structures, intergranular deformation incompatibility, or crack initiation and propagation. The calculated stress should therefore be interpreted as an effective particle-scale chemo-mechanical field and degradation precursor rather than a direct prediction of grain-boundary crack paths or cycle life. The results also provide quantitative guidance for tailoring particle morphology to manage the capacity-stress trade-off in active electrode materials.
 

\section*{CRediT authorship contribution statement}
\textbf{Rongyue Lin:}  Writing – original draft, Writing – review \& editing, Visualization, Software,  Methodology, Investigation, Formal analysis, Data curation. \textbf{Royal C. Ihuaenyi:} Writing – original draft, Writing – review \& editing, Visualization, Software, Methodology. \textbf{Juner Zhu:} Writing – review \& editing, Supervision, Resources, Project administration, Methodology, Investigation, Funding acquisition, Conceptualization. \textbf{Ruobing Bai:} Writing – original draft, Writing – review \& editing, Supervision, Resources, Project administration, Methodology, Investigation, Funding acquisition, Conceptualization.

\section* {Declaration of competing interest}
The authors declare no conflict of interest.

\section* {Acknowledgements}
This research was supported by the U.S. National Science Foundation through grants CMMI-2450006 and CMMI-2543158. R. L. acknowledges the support from Northeastern University Mechanical and Industrial Engineering Chair Fellowship.

\newpage
\appendix
\section{Chemo-mechanical contribution to Li transport in spherical particles}\label{app:analytical for sphere}


\begin{figure}[H]
    \centering
    \includegraphics[width=0.9\linewidth]{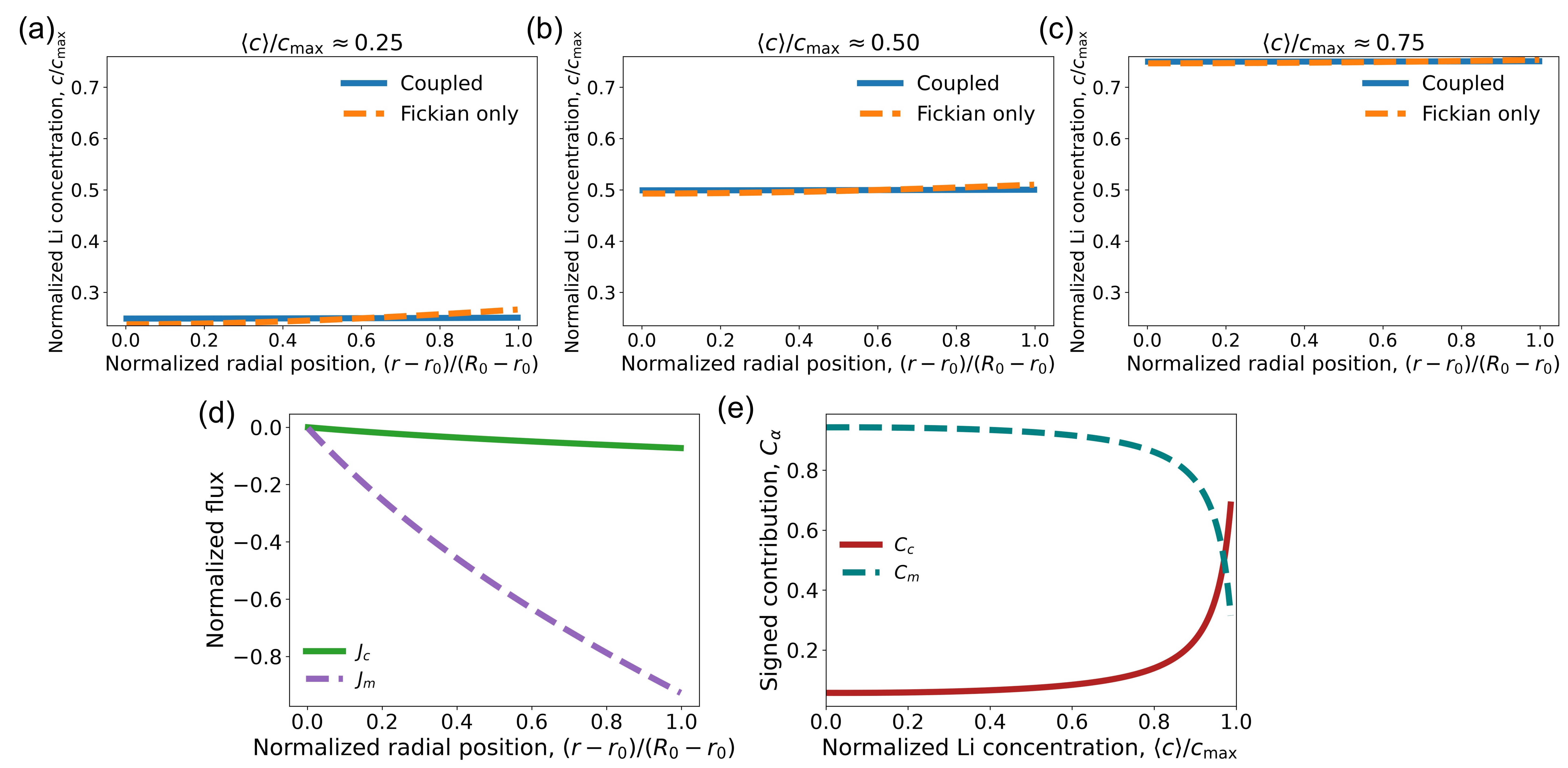}
    \caption{Assessment of chemo-mechanical coupling in Li transport in an isotropic hollow spherical particle. (a–c) Radial profiles of normalized Li concentration, \(c/c_{\max}\), predicted by the fully coupled and concentration-driven (Fickian only) models at normalized average lithiation states $\left<c\right>/c_{\max}$=0.25, 0.50, and 0.75. (d) Radial distributions of the chemical flux  \(J_{\mathrm{c}}\) and the mechanical flux \(J_{\mathrm{m}}\) at \(\left<c\right>/c_{\max}=0.5\). (e) Evolution of the global flux ratios \(C_{\mathrm{c}}\) and \(C_{\mathrm{m}}\) with normalized average lithiation, \(\left<c\right>/c_{\max}\).}
    \label{app fig:c profile}
\end{figure}


\section{Effect of particle aspect ratio}
\label{app:aspect ratio}
To isolate the effect of particle morphology on lithiation heterogeneity and chemo-mechanical transport, we perform a systematic aspect-ratio study connecting the spherical and ellipsoidal geometries considered in the main text. We define a geometric parameter \(\eta =a_{\mathrm{out}}/c_{\mathrm{out}}\), and consider \(\eta\)=1.0, 0.8, 0.6, 0.5, where \(\eta\)=1 corresponds to a sphere and \(\eta\)=0.5 corresponds to the elongated ellipsoidal geometry studied in the main text. To isolate the effect of morphology, all cases use the same isotropic material properties and the same inner-free mechanical constraint. The outer-particle volume and internal void dimensions are held constant. Because the external surface area varies with aspect ratio, the applied local insertion flux is adjusted for each geometry so that the same nominal total 1C lithiation rate is maintained.

\begin{figure}
    \centering
    \includegraphics[width=0.9\linewidth]{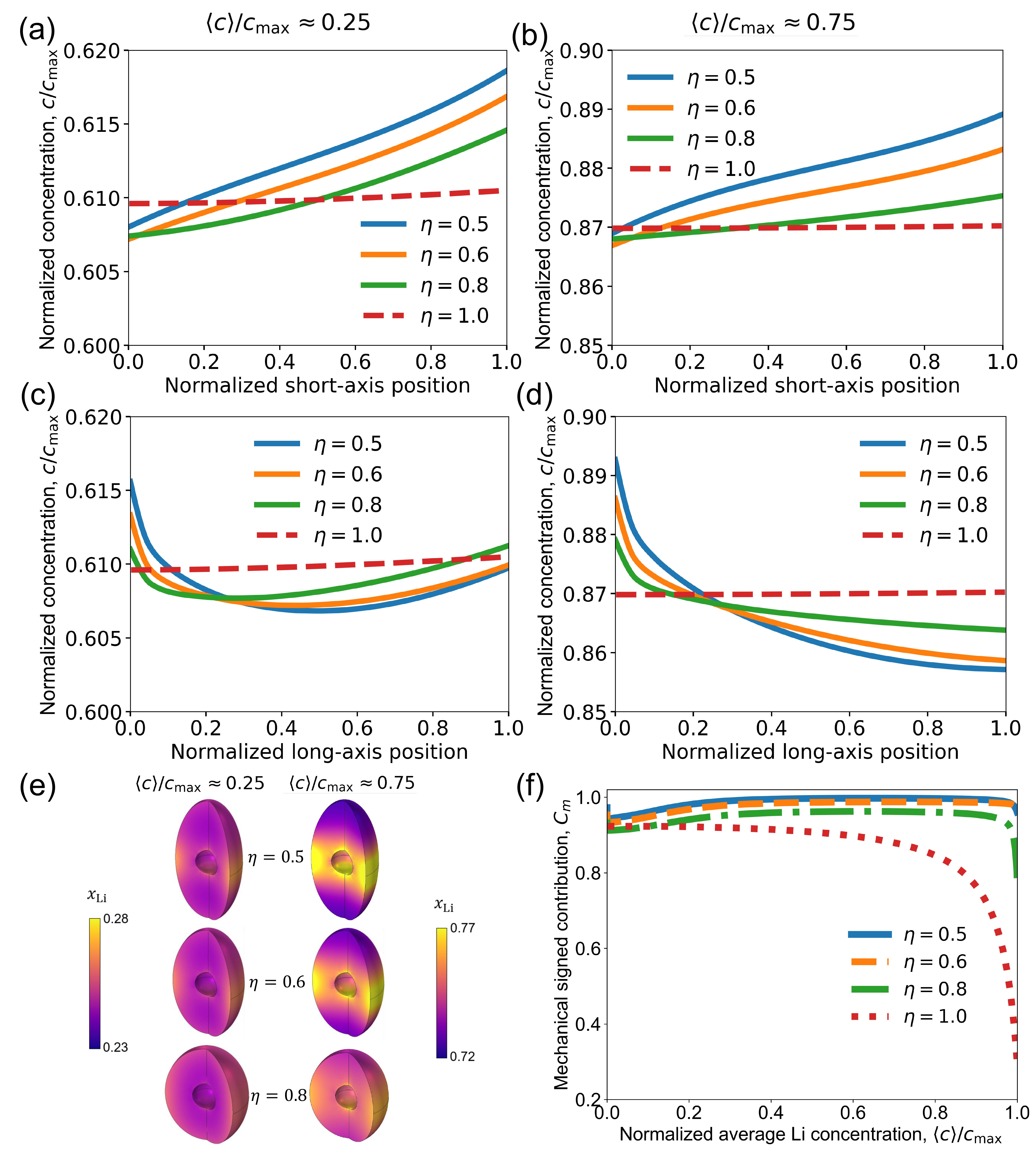}
    \caption{Effect of particle aspect ratio on lithiation heterogeneity and mechanical contribution to Li transport. (a,b) Normalized Li concentration profiles along the short axis at normalized volume-averaged lithiation states \(\left<c\right>/c_{\text{max}}\)=0.25 and 0.75, respectively. (c,d) Corresponding concentration profiles along the long axis. (e) Representative normalized Li concentration fields illustrating the progressive development of directional lithiation heterogeneity as the aspect ratio decreases. (f) Evolution of the mechanical global flux ratio \(C_{\mathrm{m}}\) during lithiation for different aspect ratios \(\eta\).}
    \label{app fig:aspect ratio}
\end{figure}

Figs.~\ref{app fig:aspect ratio}a–d show the concentration profiles along the short and long axes at representative average lithiation states \(\left<c\right>/c_{\text{max}}=0.25\) and \(0.75\). For the spherical particle, the short- and long-axis profiles coincide, as expected from spherical symmetry. As \(\eta\) decreases, the directional concentration distributions progressively depart from the spherical result. Along the short axis, the surface-to-interior concentration difference increases with increasing elongation, with the strongest gradient occurring at \(\eta=0.5\). Along the long axis, departure from the spherical geometry produces a substantially different concentration distribution, including pronounced interior minima at early lithiation and increasingly strong axial concentration gradients at later lithiation. The divergence between the short- and long-axis profiles therefore increases as the particle becomes more elongated (Fig. \ref{app fig:aspect ratio}e), demonstrating that nonspherical geometry alone can induce increasingly anisotropic lithiation even when the material properties themselves remain isotropic.

This geometrical effect is also reflected in the flux decomposition (Fig. \ref{app fig:aspect ratio}f). The spherical particle exhibits the smallest mechanical global flux ratio \(C_m\), which remains approximately 0.9 over much of lithiation before decreasing to below 0.4 near full lithiation. \(C_{\mathrm{m}}\) further increases with the decrease of \(\eta\), and reaches close to unity over much of lithiation in the most elongated particle, \(\eta=0.5\). Therefore, as the particle departs from the spherical geometry and becomes further elongated, the mechanical contribution to the chemical potential and flux becomes more significant.

\section{Analytical expressions in cylindrical particles}
\label{app:analytical for cylinders}

Under cylindrical symmetry, the radial flux in a transversely isotropic cylindrical particle is expressed as
\begin{equation}
\begin{aligned}
J_r={}&-D_{a}\frac{c_{\max}}{c_{\max}-c}\frac{\partial c}{\partial r} \\
&+\frac{D_{a} c}{RT}\left[
\left((C_{11}+C_{12})\beta_{a}+C_{13}\beta_c\right)
\frac{\partial }{\partial r}\left(\frac{\partial u_r}{\partial r}+\frac{u_r}{r}\right)
+
\left(2C_{13}\beta_{a}+C_{33}\beta_c\right)\frac{\partial \varepsilon_{zz}}{\partial r}
\right] \\
&-\frac{D_{a} c}{RT}\frac{\partial c}{\partial r}
\left[
2(C_{11}+C_{12})\beta_{a}^2
+4C_{13}\beta_{a}\beta_c
+C_{33}\beta_c^2
\right].
\end{aligned}
\label{app eqn:cylindrical flux}
\end{equation}

In the axial direction, we assume that the particle is either fully constrained in displacement (axial strain $\varepsilon_{zz}=0$, Fig.~\ref{fig: energy comparison}b left) or unconstrained (total axial force $\int_{r_0}^{R_0} \sigma_{zz}rdr = 0$, Fig.~\ref{fig: energy comparison}c left). The former leads to an axial stress of
\begin{equation}
   \sigma_{zz}
   =
   C_{13}\left(\frac{\partial u_r}{\partial r}+\frac{u_r}{r}\right)
   -
   \left(2C_{13}\beta_{a}+C_{33}\beta_c\right)\tilde{c},
   \label{app eqn:cylinder axial stress}
   \end{equation}
while the latter leads to an axial strain of
\begin{equation}
  \varepsilon_{zz}
  =
  -\frac{2C_{13}}{C_{33}}\frac{R_0u_r(R_0)-r_0u_r(r_0)}{R_0^2-r_0^2}
  +
  \frac{2\left(2C_{13}\beta_{a}+C_{33}\beta_c\right)}{(R_0^2-r_0^2)C_{33}}
  \int_{r_0}^{R_0}\tilde{c}\,r\,\text{d}r.
  \label{app eqn:cylinder axial strain}
\end{equation}

For isotropic cylindrical particles, Eqs. \ref{app eqn:cylindrical flux}-\ref{app eqn:cylinder axial strain} further reduce to
\begin{equation}
    J_r
    =
    -D\frac{c_{\max}}{c_{\max}-c}\frac{\partial c}{\partial r}
    +
    \frac{Dc}{RT}(3\lambda+2G)\beta
    \frac{\partial }{\partial r}\left(\frac{\partial u_r}{\partial r}+\frac{u_r}{r}\right)
    -
    \frac{Dc}{RT}(9\lambda+6G)\beta^2\frac{\partial c}{\partial r},
\end{equation}
\begin{equation}
    \sigma_{zz}
    =
    \lambda\left(\frac{\partial u_r}{\partial r}+\frac{u_r}{r}\right)
    -
    (3\lambda+2G)\beta\,\tilde{c},
\end{equation}
and
\begin{equation}
    \varepsilon_z
    =
    -\frac{2\lambda}{(\lambda+2G)(R_0^2-r_0^2)}
    \left(R_0u_r(R_0)-r_0u_r(r_0)\right)
    +
    \frac{2(3\lambda+2G)\beta}{(\lambda+2G)(R_0^2-r_0^2)}
    \int_{r_0}^{R_0}\tilde{c}\,r\,\text{d}r.
\end{equation}

\section{Analytical expressions in ellipsoidal particles}
\label{app:analytical for ellips}

Under cylindrical symmetry, the flux components along the $r$ and $z$ axes in a transversely isotropic ellipsoidal particle are expressed as
\begin{equation}
\begin{aligned}
J_r={}&-D_c\frac{c_{\max}}{c_{\max}-c}\frac{\partial c}{\partial r} \\
&+\frac{D_c c}{RT}\left[
\left(C_{33}\beta_c+2C_{13}\beta_{a}\right)\frac{\partial \varepsilon_{rr}}{\partial r}
+
\left(C_{13}\beta_c+(C_{11}+C_{12})\beta_{a}\right)\frac{\partial \varepsilon_{\theta\theta}}{\partial r}
+
\left(C_{13}\beta_c+(C_{11}+C_{12})\beta_{a}\right)\frac{\partial \varepsilon_{zz}}{\partial r}
\right] \\
&-\frac{D_c c}{RT}\frac{\partial c}{\partial r}
\left[
C_{33}\beta_c^2
+4C_{13}\beta_c\beta_{a}
+2(C_{11}+C_{12})\beta_{a}^2
\right],
\end{aligned}
\end{equation}
and
\begin{equation}
\begin{aligned}
J_z={}&-D_{a}\frac{c_{\max}}{c_{\max}-c}\frac{\partial c}{\partial z} \\
&+\frac{D_{a} c}{RT}\left[
\left(C_{33}\beta_c+2C_{13}\beta_{a}\right)\frac{\partial \varepsilon_{rr}}{\partial z}
+
\left(C_{13}\beta_c+(C_{11}+C_{12})\beta_{a}\right)\frac{\partial \varepsilon_{\theta\theta}}{\partial z}
+
\left(C_{13}\beta_c+(C_{11}+C_{12})\beta_{a}\right)\frac{\partial \varepsilon_{zz}}{\partial z}
\right] \\
&-\frac{D_{a} c}{RT}\frac{\partial c}{\partial z}
\left[
C_{33}\beta_c^2
+4C_{13}\beta_c\beta_{a}
+2(C_{11}+C_{12})\beta_{a}^2
\right].
\end{aligned}
\end{equation}
In the isotropic particle, the flux reduces to
\begin{equation}
    \boldsymbol{J}
    =
    -\left[
    D\frac{c_{\max}}{c_{\max}-c}
    +
    \frac{Dc}{RT}(9\lambda+6G)\beta^2
    \right]\nabla c
    +
    \frac{Dc}{RT}(3\lambda+2G)\beta\,\nabla \mathrm{tr}(\boldsymbol{\varepsilon}).
\label{app eqn:iso ellip}
\end{equation}

When considering a transversely isotropic ellipsoidal particle with an arbitrary crystal orientation of angle $\alpha$ in Section \ref{sec:opt}, the axisymmetric flux is expressed as
\begin{equation}
\boldsymbol J
=
-\frac{1}{RT}\,\boldsymbol D(\alpha)\Bigg[
\frac{RT\,c_{\max}}{c_{\max}-c}\,\nabla c
- c\,\nabla\!\big(\boldsymbol B(\alpha):\boldsymbol\varepsilon\big)
+ c\,\big(\boldsymbol B(\alpha):\boldsymbol\beta(\alpha)\big)\,\nabla c
\Bigg]
\label{eq:oriented flux}
\end{equation}
where $\boldsymbol J=[J_r,\;J_z]^T$, $\nabla c=[\partial_r c,\;\partial_z c]^T$, $\boldsymbol\varepsilon$ is the strain tensor, and $\boldsymbol{D}(\alpha)$ is the orientation-dependent diffusivity tensor projected onto the $(r,z)$ plane. The second-rank tensor $\boldsymbol B(\alpha)$ is defined as
\begin{equation}
    B_{ij}(\alpha)=C_{ijkl}(\alpha)\,\beta_{kl}(\alpha)
\end{equation}
such that $\boldsymbol B(\alpha):\boldsymbol\varepsilon=B_{ij}(\alpha)\varepsilon_{ij}$ and $\boldsymbol B(\alpha):\boldsymbol\beta(\alpha)=C_{ijkl}(\alpha)\beta_{ij}(\alpha)\beta_{kl}(\alpha)$ are orientation-dependent scalars.

\newpage
 \bibliographystyle{elsarticle-harv}
 \bibliography{cas-refs}
 
\end{document}